\documentclass[a4paper,fleqn]{cas-sc}

\usepackage[numbers]{natbib}

\def\tsc#1{\csdef{#1}{\textsc{\lowercase{#1}}\xspace}}
\tsc{WGM}
\tsc{QE}
\tsc{EP}
\tsc{PMS}
\tsc{BEC}
\tsc{DE}

\usepackage{bm}
\usepackage{algorithm}
\usepackage{algpseudocode}

\newcommand{\diff}{\,\mathrm d}

\newcommand{\PrNum}{\mathrm{Pr}}
\newcommand{\ReNum}{\mathrm{Re}}
\newcommand{\MaNum}{\mathrm{Ma}}
\newcommand{\KnNum}{\mathrm{Kn}}

\newcommand{\calF}{\mathcal{F}}

\newcommand{\calR}{\mathcal{R}}

\newcommand{\sgn}{\operatorname{sgn}}

\newcommand{\safeincludegraphics}[2][]{%
\IfFileExists{#2}{\includegraphics[#1]{#2}}{\fbox{\parbox{0.82\linewidth}{Missing figure file: \texttt{\detokenize{#2}}}}}%
}

\begin{document}
\let\WriteBookmarks\relax
\def\floatpagepagefraction{1}
\def\textpagefraction{.001}
\shorttitle{Diffuse-Interface Kinetic Model}
\shortauthors{C. Qin et~al.}

\title [mode = title]{A flux-balanced diffuse-interface kinetic model for rarefied gas-solid boundaries with arbitrary absorption probability}                      
\author[1,2,3]{Chujin Qin}[
                           orcid=000-0003-2690-9514,
                           ]

\credit{Conceptualization, , Writing - Original draft}

\author[1,2,3]{Dongke Sun}[
               orcid=0000-0002-3521-5381,
               ]
\cormark[1]
\ead{dksun@seu.edu.cn}
\credit{Supervision, Data curation}

\affiliation[1]{organization={Key Laboratory of Structure and Thermal Protection of High Speed Aircraft, Ministry of Education},
                country={P. R. China }}

\affiliation[2]{organization={Jiangsu Key Laboratory for Biomaterials and Devices, Southeast University},
                addressline={Dingjiaqiao 87}, 
                postcode={210009}, 
                city={Nanjing},
                country={P. R. China}}

\affiliation[3]{organization={School of Mechanical Engineering, Southeast University},
                city={Nanjing},
                postcode={211189}, 
                country={P. R. China}}
                
\cortext[cor1]{Corresponding author}


\begin{abstract}
    Rarefied gas--solid interactions determine mechanical loads, heat transfer, and material exchange in microscale devices and low-pressure processings.
    Modeling these interactions becomes challenging when gas non-equilibrium effect and complex moving boundaries must be resolved together.
    We develop a diffuse-interface kinetic model (DIKM) that describes rarefied gas transport and gas--solid interactions within one kinetic formulation on a single fixed Eulerian mesh.
    A smooth order parameter represents the solid geometry, while a localized molecular flux operator incorporates boundary interactions directly into the kinetic equation.
    The operator distinguishes incident and reflected distributions in the interface region and specifies their mass, momentum, and energy exchange, including prescribed surface absorption.
    Matched asymptotic analysis shows that the diffuse formulation recovers the corresponding partially absorbing diffuse-reflection boundary condition for smooth stationary or moving surfaces in the sharp-interface limit.
    The discretization uses Strang-splitting, combining an order-parameter-weighted discrete unified gas kinetic scheme for transport substep with a positivity-preserving update of the interfacial source.
    Interfacial mass and heat exchange is evaluated from the source-induced distribution increments, ensuring consistent gas--solid coupling.
    Benchmarks over planar and curved, stationary and moving boundaries reproduce the rarefaction-dependent velocity slip and wall heat transfer, absorption-driven flow, and the feedback between gas forces and solid translation and rotation.
    In the final simulations, the order parameter is evolved according to a Cahn--Hilliard equation with degenerate mobility, and the resulting thin-film growth exhibits distinct surface profiles at different vapor incidence angles.
    By connecting implicit geometry to molecular surface exchange, the DIKM provides a basis for coupling rarefied gas dynamics to solid motion and material evolution.
\end{abstract}


\begin{highlights}
    \item A flux-balanced \emph{diffuse-interface kinetic model} is proposed for gas-solid interactions in rarefied flows.
    \item Its sharp-interface limit corresponds to the partially absorbing diffuse-reflection boundary condition.
    \item Simulations demonstrate the model's ability to capture non-equilibrium effects at stationary and moving gas-solid boundaries.
\end{highlights}


\maketitle

\section{Introduction}
\label{secIntroduction}

Interfacial heat and mass transfer between a rarefied gas and an evolving solid surface underlies a range of material processing and extreme-environment flows, including physical vapor deposition (PVD) in vacuum chambers \cite{Bobzin2013}, thermal-protection coating ablation during atmospheric reentry \cite{Mansour2024}, and gasification-induced pore evolution in combustion systems \cite{WangFY2001}.
At such surfaces, molecular absorption into  or release from the solid exchange mass, momentum, and energy between the two phases.
Growth or recession rates, surface morphology, and the associated thermal and mechanical loads are therefore governed by the coupling between rarefied gas transport and interface evolution.
The associated modeling challenges are both kinetic and geometric.
As the molecular mean free path becomes comparable to relevant flow length scales, a Navier--Stokes--Fourier (NSF) description requires kinetic corrections and eventually ceases to resolve non-equilibrium transport.
A molecular description is then needed for velocity slip, temperature jump, Knudsen layers, and the interface fluxes that determine surface exchange.
The computational problem is therefore to resolve rarefied gas transport while maintaining a physically consistent molecular boundary condition on an evolving surface.

Computational kinetic methods provide several routes to the transport problem.
The direct simulation Monte Carlo (DSMC) method models molecular motion and collisions stochastically \cite{Bird1998}, but statistical noise and kinetic-scale resolution can make low-speed and near-continuum calculations costly.
Deterministic methods discretize the molecular distribution in physical and velocity space, solving the Boltzmann equation \cite{WuL2013JCP,LiuC2016JCP} or a kinetic model equation.
For discrete velocity methods (DVM) that separate transport and collision without preserving their coupled dynamics, accurate near-continuum solutions require resolution of molecular scales.
The unified gas-kinetic scheme (UGKS) \cite{XuK2010JCP} and discrete unified gas kinetic scheme (DUGKS) \cite{GuoZL2013DUGKS,GuoZL2015DUGKS,QinCJ2025AMM} address this difficulty through coupled transport--collision fluxes.
Related developments include the unified gas-kinetic wave-particle (UGKWP) method combining wave and particle representations \cite{LiuC2020JCP}, and a general synthetic iterative scheme (GSIS) accelerating convergence through coupled macroscopic equations \cite{ZhuLH2021JCP}.
The unified preserving framework further characterizes the hydrodynamic and kinetic asymptotics of discretization schemes \cite{GuoZL2023UP}.
These developments improve bulk transport calculations, but consistency of a regularized gas--solid boundary remains a separate requirement \cite{XinZY2025PRE}.

Kinetic solvers have been combined with several treatments of moving solid boundaries.
Moving-grid and arbitrary Lagrangian--Eulerian (ALE) approaches adapt the spatial discretization to boundary motion \cite{Tiwari2022JCP,ZengJN2025JCP}.
Cartesian cut-cell methods retain a background grid and reconstruct the solid surface within intersected cells, allowing conservative treatment of molecular transport around moving obstacles \cite{Dechriste2016JCP}.
Immersed boundary methods also use a background grid, with boundary information supplied through an explicit surface representation and local reconstruction or forcing \cite{TaoS2018JCP,HeQ2024IJHMT}.
He et al.\ \cite{HeQ2024IJHMT} combined DUGKS with an immersed boundary treatment for thermal rarefied flows with moving boundaries.
Xu et al.\ \cite{XuD2026JCP} developed a kinetic immersed boundary formulation in which a Galilean-invariant source is related to distribution-function discontinuities across the boundary.
Their treatment reconstructs distributions on both sides of the immersed boundary layer and evaluates surface stress and heat flux.
In addition, for complex geometries on Cartesian meshes, Filbet and Yang \cite{Filbet2013JCP} imposed kinetic boundary conditions through an inverse Lax--Wendroff reconstruction at ghost points outside the gas domain.
These methods provide established routes to resolve kinetic wall interactions.
Nevertheless, maintaining surface representations and geometric reconstructions remains a computational task when deposition or recession changes the boundary shape.

Diffuse-interface methods describe geometry implicitly through a smooth order parameter defined on a fixed mesh \cite{Anderson1998}.
A finite transition layer replaces the sharp gas--solid boundary, so the boundary motion and possible topological changes can be represented through order-parameter evolution.
For this representation to impose physical boundary conditions, distributed terms within the layer must recover the corresponding sharp-interface conditions as the layer thickness vanishes.
Li et al.\ \cite{LiX2009} developed a diffuse-interface approach for second-order equations with prescribed boundary values or fluxes, while Yu et al.\ \cite{YuH2012} extended the formulation to general boundary conditions and surface reaction--diffusion problems.
For viscous compressible flow, Boyd et al.\ \cite{Boyd2026JCP} represented interfacial mass, momentum and energy fluxes by distributed sources.
Asymptotic studies of diffuse-interface methods show that the recovered boundary condition and its accuracy depend on the regularization \cite{Lervag2014,YuF2020JCP}.
At the kinetic level, a gas--surface boundary law relates incident and outgoing molecular distributions \cite{Brull2014KRM}, which are not determined by a finite set of macroscopic values or fluxes.

Implicit gas--solid representations have also been introduced directly into kinetic equations.
Sato et al.\ \cite{Sato2019JCP} extended a kinetic model over gas and solid regions for topology optimization, approximating diffuse reflection through material-dependent equilibrium velocity, temperature, and collision frequency.
Yuan and Wu \cite{YuanRF2024JCP} refined this fictitious-porosity approach by separating the material interpolation of the target velocity and temperature from that of the relaxation time, reducing undesirable intermediate-density regions in rarefied-flow optimization.
These studies establish implicit kinetic boundary treatments for steady design optimization.
For the relaxation-based models, boundary recovery depends on the material interpolation and the relaxation scale assigned to the solid, whereas recovery for a binary material distribution does not by itself establish the sharp-interface limit of a smooth moving layer.
Guan and Yamada \cite{GuanKW2024JCP} extended DVM to mixed gas--solid domains through material-dependent convective fluxes and reflection sources that account separately for incident losses and reflected gains.
It recovers the underlying discrete wall treatment for binary gas--solid distributions.
An evolving smooth interface requires a further connection between the molecular boundary law, the distributed exchange, and the evolution of the material region.
Therefore, a diffuse kinetic formulation is required that recovers a prescribed molecular boundary law in sharp-interface limit and uses interfacial exchange to determine gas-phase balances and material evolution.
Establishing this formulation would allow rarefied transport, wall loads, and deposition-driven geometry changes to be treated within a consistent description on a fixed mesh.

Surface absorption makes partition of incident, absorbed and emitted molecules essential to both gas transport and material accumulation.
Kinetic studies have established the influence of absorption on rarefied mixture flows \cite{Frezzotti2012}, and recent multiscale simulations have treated absorbing walls in gas mixtures \cite{LiW2025PoF}.
An absorption probability specifies the fraction of incident molecular flux removed from the gas; the remaining fraction is returned according to a reflection law.
Here, non-absorbed molecules are assumed to be diffusely reflected with a Maxwellian distribution at the wall velocity and temperature.
Incident and reflected populations occupy opposite halves of velocity space, distinguished by the sign of molecular normal velocity relative to the wall.
The boundary-modeling task is to preserve this relation within a finite transition layer and recover the prescribed absorption--diffuse-reflection law as the layer thickness vanishes.
The resulting mass exchange must reduce to zero for a non-absorbing wall and to the prescribed sink for partial or complete absorption.

In this work, we develop a flux-balanced diffuse-interface kinetic model (DIKM) that couples rarefied gas transport and molecular surface exchange on a fixed Eulerian mesh, including moving boundaries and prescribed absorption.
The model represents boundary interactions through a localized molecular flux operator and recovers the prescribed sharp-interface law through matched asymptotic analysis for smooth, kinematically compatible interfaces.
A localized molecular flux operator is introduced to incorporate incident removal and diffuse reflection, with its velocity moments determining interfacial mass, momentum and energy transfer.
The remainder of the paper is organized as follows.
Section~\ref{secMethod} presents the gas kinetic equation and boundary conditions, the DIKM formulation and its sharp-interface limit.
Section~\ref{secSolveMethod} describes the numerical discretization and interfacial exchange.
Section~\ref{secResult} reports benchmark comparisons and the thin-film growth example.
Section~\ref{secConclusion} summarizes the conclusions and perspectives.

\section{Diffuse-interface kinetic model}
\label{secMethod}

The starting point is the gas kinetic equation governing a rarefied gas bounded by a sharp moving wall subject to an absorbing diffuse-reflection law.
A smooth gas indicator and a regularized surface measure are then introduced to distribute the interfacial fluxes over a finite transition layer, where they act as a local operator on the weighted kinetic equation and thereby define the DIKM.
The moments of this operator quantify the interfacial mass, momentum, and energy exchange.
Matched asymptotics establishes the recovery of the prescribed sharp-interface law.

\subsection{Gas kinetic model}
\label{subsec:gas_model}

Let $\Omega_g(t)\subset\Omega\subset\mathbb R^D$ be the gas region inside a fixed, larger and regular computational domain $\Omega$, where $D$ is the number of resolved spatial dimensions.
The gas--solid interface is denoted by $\Gamma(t)$.
External computational boundaries $\partial \Omega$ are treated separately.
The density distribution function $f(\bm x,\bm\xi,\bm\eta,\bm\zeta,t)$ depends on position $\bm x$, time $t$, resolved molecular velocity $\bm\xi\in\mathbb R^D$, unresolved translational velocity $\bm\eta\in\mathbb R^{3-D}$, and internal coordinates $\bm\zeta\in\mathbb R^K$.
The $K$ internal degrees of freedom are assumed to equilibrate at the translational temperature.
In the absence of external body forces, the Shakhov equation is adopted for gas bulk transport
\begin{equation}
    \label{eqShakhovEq}
    \dfrac{\partial f}{\partial t}+\bm\xi\cdot\nabla f=Q^{\rm S}(f),
    \qquad Q^{\rm S}(f)=-\frac{f-f^{\rm S}}{\tau}.
\end{equation}
Here $\tau$ is the relaxation time.
The Shakhov distribution and Maxwellian equilibrium state are
\begin{align}
    f^{\rm S}&=f^{\rm M}\left[1+(1-\PrNum)
    \frac{\bm c\cdot\bm q}{5pRT}
    \left(\frac{c^2+\eta^2}{RT}-5\right)\right],
    \label{eqShakovDF}
    \\
    f^{\rm M}&=\frac{\rho}{(2\pi RT)^{(3+K)/2}}
    \exp\!\left[-\frac{c^2+\eta^2+\zeta^2}{2RT}\right],
    \label{eqMaxwellDF}
\end{align}
where $\bm c=\bm\xi-\bm u$, $c^2=\bm c\cdot\bm c$, and analogous squared norms are used for the other velocities.
The quantities $\rho$, $\bm u$, $T$, $p=\rho RT$, and $\bm q$ are the gas density, velocity, temperature, pressure, and heat flux, respectively.
$R$ is the specific gas constant and $\PrNum$ is the prescribed Prandtl number.

With $\diff\bm\varXi=\diff\bm\xi\diff\bm\eta\diff\bm\zeta$, the conservative moments are
\begin{equation}
    \rho=\int f\diff\bm\varXi,\qquad
    \rho\bm u=\int\bm\xi f\diff\bm\varXi,\qquad
    \rho E=\frac12\int(\xi^2+\eta^2+\zeta^2)f\diff\bm\varXi.
    \label{eqDF2Rho}
\end{equation}
The specific total energy is $E=|\bm u|^2/2+c_vT$, with $c_v=(3+K)R/2$ and heat-capacity ratio $\gamma=(K+5)/(K+3)$.
The pressure tensor $\bm P$ and heat flux $\bm q$ are
\begin{equation}
    \bm P=\int\bm c\bm c f\diff\bm\varXi=p\bm I+\bm\Pi,
    \qquad
    \bm q=\frac12\int\bm c(c^2+\eta^2+\zeta^2)f\diff\bm\varXi.
    \label{eqDF2tau}
\end{equation}
Here $\bm I$ is the identity tensor, $\bm\Pi$ is the non-equilibrium part of the pressure tensor.
The Shakhov collision term conserves the moments in Eq.~\eqref{eqDF2Rho} and relaxes the heat flux at rate $\PrNum/\tau$.

The relaxation time is related to the dynamic viscosity by
\[
    \tau=\mu/p,
\]
with the viscosity modeled by a temperature-dependent power law:
\begin{equation}
    \mu=\mu_0\left(\frac{T}{T_0}\right)^\omega,
    \label{eq:viscosity_model}
\end{equation}
where $\mu_0$ is the viscosity at reference temperature $T_0$ and $\omega$ is the viscosity exponent.
The hard-sphere (HS) molecule model has $\omega=0.5$, and the variable-hard-sphere (VHS) model considered here uses $\omega=0.81$.
Flow conditions are usually characterized by certain dimensionless numbers, including Knudsen number $\KnNum=\lambda/L_c$, Reynolds number $\ReNum=\rho_0U_cL_c/\mu_0$, and Mach number $\MaNum=U_c/\sqrt{\gamma RT_0}$, where $L_c$ and $U_c$ are reference length and velocity, $\rho_0$ is the reference density, and $\lambda$ is the reference molecular mean free path given by
\begin{equation}
    \lambda = \dfrac{2(5-2\omega)(7-2\omega)}{15\pi} \dfrac{\mu_0}{\rho_0}\sqrt{\dfrac{\pi}{2RT_0}}.
\end{equation}

\subsection{Absorbing diffuse-reflection boundary condition}
\label{subsec:sharp_bc}

The sharp-interface boundary condition in this subsection specifies the flux balance that the diffuse formulation is required to reproduce.
Let the interfacial normal vector $\bm n_\Gamma (\bm s, t)$ point from the solid into the gas, and let $\bm u_\Gamma (\bm s, t)$ and $T_\Gamma (\bm s, t)$ be the wall velocity and temperature, with $\bm s\in\Gamma$.
The relative normal molecular velocity is
\begin{equation}
    c_{n}=(\bm\xi-\bm u_\Gamma)\cdot\bm n_\Gamma.
    \label{eq:normal_velocity}
\end{equation}
Molecules incident on the wall have $c_{n}<0$, whereas those emitted from it have $c_{n}>0$.
For absorption probability $\beta\in[0,1]$, independent of molecular velocity, the emitted distribution is
\begin{equation}
    f^+=\rho_\Gamma\hat f_\Gamma^{\rm M},\qquad c_{n}>0,
    \label{eq:sharp_emitted_distribution}
\end{equation}
where the unit-density wall Maxwellian is
\begin{equation}
    \hat f_\Gamma^{\rm M}
    =\frac{1}{(2\pi RT_\Gamma)^{(3+K)/2}}
    \exp\left[-\dfrac{|\bm\xi-\bm u_\Gamma|^2+\eta^2+\zeta^2}{2RT_\Gamma}\right]
    .
    \label{eq:wall_maxwellian}
\end{equation}
All non-absorbed molecules are assumed to be diffusely reflected and thermalized at $T_\Gamma$.
The density $\rho_\Gamma$ follows from the balance among the incident, emitted, and absorbed fluxes.
If $f^-$ denotes the incident distribution, returning only its non-absorbed fraction gives
\begin{equation}
    \label{eqRhoGamma}
    \rho_\Gamma=
    \frac{(1-\beta)\displaystyle\int_{c_{n}<0}|c_{n}|f^-\diff\bm\varXi}
    {\displaystyle\int_{c_{n}>0}c_{n}\hat f_\Gamma^{\rm M}\diff\bm\varXi},
\end{equation}
where the denominator equals $\sqrt{RT_\Gamma/2\pi}$.
The absorbed flux is
\begin{equation}
    J_{{\rm abs}}=\beta\int_{c_{n}<0}|c_{n}|f^-\diff\bm\varXi.
    \label{eq:sharp_absorbed_flux}
\end{equation}
For $\beta=0$, Eq.~\eqref{eq:sharp_absorbed_flux} gives zero net mass flux relative to the wall, or $\rho(\bm u-\bm u_\Gamma)\cdot\bm n_\Gamma=0$ at the sharp boundary, and Eq.~\eqref{eqRhoGamma} reduces to non-penetration condition.
For $\beta=1$, $\rho_\Gamma=0$ and no molecules are re-emitted.
Thus the practical implementation preserves the velocity half-space distinction and the prescribed partition of incident mass between absorption and reflection.

\subsection{Diffuse-interface representation and regularization}
\label{subsec:diffuse_regularization}

\begin{figure}[!htp]
    \centering
    \includegraphics[width=0.8\textwidth]{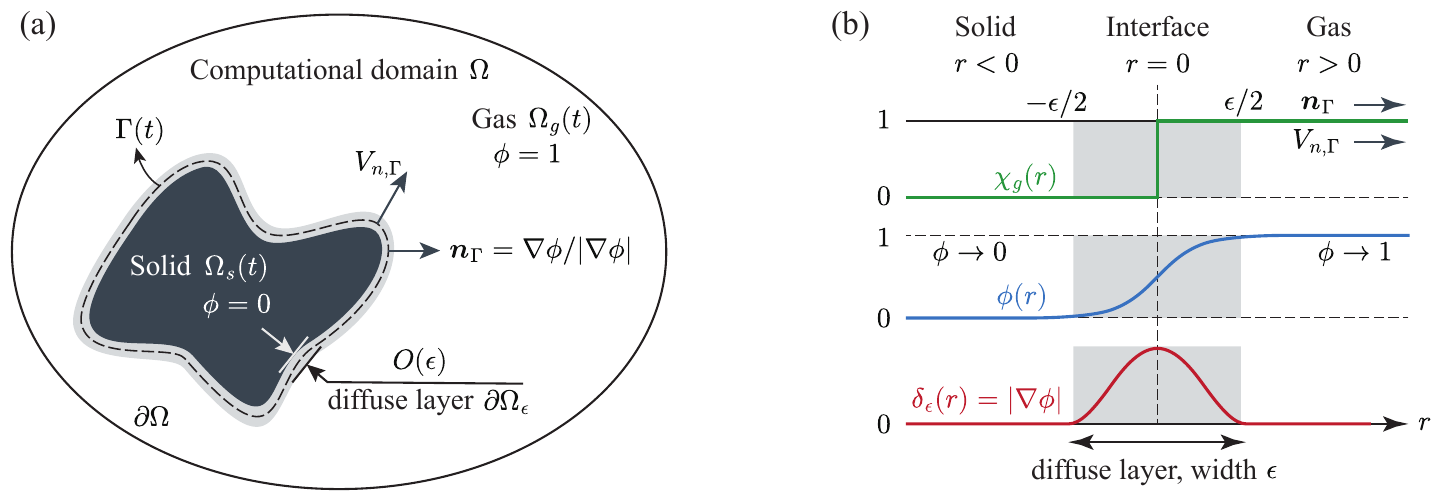}
    \caption{
    Schematic diagram of diffuse-interface representation of
    (a) computational domain and 
    (b) interface profiles in interfacial local coordinate.
    }
    \label{figDiffuseInterface}
\end{figure}

The diffuse-interface computational configuration is sketched in Fig.~\ref{figDiffuseInterface}.
Introduce the characteristic function of the moving gas domain,
\begin{equation}
    \chi_g(\bm x,t)=
    \begin{cases}
        1, & \bm x\in\Omega_g(t),\\
        0, & \bm x\in\Omega\setminus\Omega_g(t).
    \end{cases}
    \label{eq:gas_characteristic}
\end{equation}
We represent the gas region by a Lipschitz-continuous dimensionless diffuse indicator $\phi(\bm x,t)\in[0,1]$, also referred to here as an order parameter.
It provides a smooth approximation to $\chi_g(\bm x,t)$: $\phi\to1$ in the gas and $\phi\to0$ in the solid as the interface thickness tends to zero, away from $\Gamma(t)$.
We adopt the hyperbolic-tangent profile
\begin{equation}
    \phi(\bm x,t) = \dfrac12\left[1 + \tanh\left(\dfrac{3r(\bm x, t)}{\epsilon}\right)\right],
    \label{eqInitPhi}
\end{equation}
where $r(\bm x,t)$ is the signed distance to $\Gamma(t)$, positive in the gas and negative in the solid, with $|\nabla r|=1$ in a smooth tubular neighborhood of the interface.
The geometric interface is the level set $r=0$, equivalently $\phi=1/2$.
The parameter $\epsilon>0$ sets the transition scale and serves as a nominal interface width, with $\phi\approx0.05$ and $0.95$ at $r=-\epsilon/2$ and $\epsilon/2$, respectively.
This construction also accommodates geometry supplied as a segmented digital image: identify the gas--solid boundary, compute its signed-distance field, and apply Eq.~\eqref{eqInitPhi} at the grid points.
Alternatively, an auxiliary phase-field equation, such as a Cahn--Hilliard equation, may supply the phase field, provided that its range, transition normalization, and motion are consistent with the diffuse-interface profile defined by the present model.

Since $\phi$ increases from the solid to the gas, the unit normal pointing into the gas is defined by
\begin{equation}
    \label{eq:diffuse_normal}
    \bm n_\epsilon=\frac{\nabla\phi}{|\nabla\phi|}.
\end{equation}
Where the interface and indicator are sufficiently smooth, the corresponding curvature is
\begin{equation}
    \kappa_\epsilon=\nabla\cdot\bm n_\epsilon,
    \label{eq:diffuse_curvature}
\end{equation}
where $\kappa_\epsilon$ is the signed curvature of the indicator level sets (the sum of the principal curvatures in three dimensions), evaluated on $\phi=1/2$ for the geometric interface.
With the order parameter specifying the interface geometry, the molecular fluxes prescribed on $\Gamma(t)$ are also represented within the transition layer.
We introduce a nonnegative regularized surface delta function $\delta_\epsilon$, understood as a smooth approximation to the surface Dirac measure $\delta_\Gamma$ supported on $\Gamma(t)$, to convert these surface fluxes into volumetric source terms.
The regularization distributes the boundary interaction across the layer, with a unit integral in the normal direction ensuring that the prescribed flux per unit interfacial area is preserved.
Among the available surface regularizations, we adopt a simple form based on the coarea interpretation \cite{XuXP2026}
\begin{equation}
    \delta_\epsilon=|\nabla\phi|,
    \qquad
    \int_{-\infty}^{\infty}\delta_\epsilon(r)\diff r=1.
    \label{eq:diffuse_geometry}
\end{equation}
For the signed-distance profile in Eq.~\eqref{eqInitPhi}, the local normal weight is
\begin{equation}
    \delta_\epsilon(r)
    =\frac{6}{\epsilon}\phi(1-\phi).
    \label{eq:surface_delta_profile}
\end{equation}
Consequently, decreasing $\epsilon$ narrows the interaction layer and increases its peak weight as $\epsilon^{-1}$ without changing the normal integral.
The quantity $\delta_\epsilon$ has units of inverse length, so multiplying a flux per unit surface area by $\delta_\epsilon$ gives a source per unit volume.
The diffuse normal $\bm n_\epsilon$ is needed only where $\delta_\epsilon>0$, and numerical cutoff treatment of the small solid-side tails is described in Section~\ref{subsec:regularized_split}.
For a smooth interface contained in $\Omega$, let $a$ be a smooth surface quantity and $a_\epsilon$ its bounded, smooth extension, constant along local normal lines.
The corresponding volume integral approaches the surface integral \cite{Schmidt2022PoF}:
\begin{equation}
    \int_\Omega a_\epsilon \delta_\epsilon\diff V
    \longrightarrow\int_\Gamma a\diff A
    \quad\text{as }\epsilon\to0.
    \label{eq:surface_measure}
\end{equation}
Here $\diff V$ and $\diff A$ denote volume and surface measures in $D$ dimensions.
This is the distributional convergence of $\delta_\epsilon$ to the surface delta $\delta_\Gamma$.
To see the meaning of Eq.~\eqref{eq:surface_measure}, use local normal coordinates $\bm x=\bm s+r\bm n_\Gamma(\bm s)$.
The volume element is $\diff V=J(\bm s,r)\diff r\diff A$, where $J(\bm s,0)=1$ and $J\to1$ as $r\to0$.
As $\epsilon\to0$, the normalized weight concentrates at $r=0$, and integration through the layer therefore leaves $a(\bm s)$ per unit interfacial area.
For $a=1$, the volume integral estimates the total interfacial area (curve length in two dimensions).
For a surface mass flux, it gives the corresponding total mass-transfer rate.

The diffuse wall fields $\bm u_\epsilon$, $T_\epsilon$, and $\rho_\epsilon$ extend their sharp-interface counterparts into the diffuse layer.
The relative normal molecular velocity is thus given by
\begin{equation}
    c_{n}=(\bm\xi-\bm u_\epsilon)\cdot\bm n_\epsilon.
    \label{eq:app_phase}
\end{equation}
For a moving interface, the evolution of the phase field is also required to agree with the wall velocity used in the relative molecular flux.
We measure this kinematic compatibility by the residual
\begin{equation}
    E_\phi=\partial_t\phi+\bm u_\epsilon\cdot\nabla\phi.
    \label{eq:indicator_kinematic_residual}
\end{equation}
For a stationary phase field with a stationary wall, or for a fixed-width interface exactly advected by a rigid-body velocity, $E_\phi=0$.
Profile relaxation, reinitialization, or a mismatch between the geometric and wall normal velocities can instead produce a nonzero residual.
The residual $E_\phi$ measures the departure of the phase field evolution from advection by the wall velocity.
Its contribution to the kinetic balance is explained explicitly later.

\subsection{Kinetic formulation and interfacial exchange}
\label{subsec:dikm}

The boundary operator is required to remove the incident molecular flux and to locally supply the diffusely reflected flux emitted from the wall.
Multiplication by the regularized surface measure extends the kinetic equation to a larger and regular computational domain $\Omega$.
With the detailed derivation given in Appendix~\ref{secAppendixA}, the diffuse-interface kinetic model (DIKM) is formulated as
\begin{equation}
    \label{eqDIKM}
    \dfrac{\partial (\phi f)}{\partial t}
    +\bm\xi\cdot\nabla(\phi f)
    =\phi Q^{\rm S}(f)+\delta_\epsilon\mathcal R,
\end{equation}
with the local boundary-flux operator
\begin{equation}
    \mathcal  R
    =-\chi^-|c_{n}|f
    +\chi^+ c_{n}\rho_\epsilon\hat f_\epsilon^{\rm M},
    \label{eq:local_boundary_operator}
\end{equation}
where $\chi^-=1$ for $c_{n}<0$ and zero otherwise, and $\chi^+=1$ for $c_{n}>0$ and zero otherwise.
The distribution $\hat f_\epsilon^{\rm M}$ is Eq.~\eqref{eq:wall_maxwellian} evaluated with the diffuse wall fields.
The emitted density $\rho_\epsilon$ is determined from the local incident distribution and non-absorbed fraction
\begin{equation}
    \rho_\epsilon=
    \frac{(1-\beta)\displaystyle\int_{c_{n}<0}|c_{n}|f\diff\bm\varXi}
    {\displaystyle\int_{c_{n}>0}c_{n}\hat f_\epsilon^{\rm M}\diff\bm\varXi}.
    \label{eq:rho_gamma_local}
\end{equation}
This ensures that the net mass sink equals the absorbed fraction of the incident flux, including the non-absorbing and fully absorbing limits.
The limiting cases $\beta = 0$ and $\beta = 1$ correspond to non-penetration and fully absorbing boundary conditions, respectively.
The explicit half-space flux balance, together with its spatial distribution across the diffuse-interface layer, connects the diffuse representation to the prescribed molecular boundary condition in Section~\ref{subsec:sharp_bc}.
The diffuse-interface boundary condition is formulated directly at the kinetic level and incorporated into the kinetic equation, allowing it to accommodate arbitrary boundary geometries without requiring a body-fitted mesh or explicit interface tracking.

To further interpret the transport and boundary-flux terms in Eq.~\eqref{eqDIKM}, apply the product rule with $\nabla\phi=\delta_\epsilon\bm n_\epsilon$ and the kinematic residual defined in Eq.~\eqref{eq:indicator_kinematic_residual}:
\begin{equation}
    \partial_t(\phi f)+\bm\xi\cdot\nabla(\phi f)
    =\phi(\partial_t f+\bm\xi\cdot\nabla f)
    +\delta_\epsilon c_n f+fE_\phi.
    \label{eq:weighted_transport_identity}
\end{equation}
Combining Eqs.~\eqref{eq:weighted_transport_identity} and \eqref{eqDIKM} gives
\begin{equation}
    \phi\left[\partial_t f+\bm\xi\cdot\nabla f-Q^{\rm S}(f)\right]
    =\delta_\epsilon(\mathcal R-c_n f)-fE_\phi.
    \label{eq:unweighted_diffuse_balance}
\end{equation}
For incident velocities, $\mathcal R=c_n f$, so the surface terms cancel.
When $E_\phi=0$, the incident gas distribution satisfies the bulk kinetic equation wherever $\phi>0$.
For emitted velocities, $\mathcal R-c_n f=c_n(\rho_\epsilon\hat f_\epsilon^{\rm M}-f)$ supplies the wall-emitted state.
A nonzero $E_\phi$ also modifies the incident balance, so an auxiliary evolution of $\phi$ is admissible only when it remains consistent with the wall kinematics.
The later sharp-interface analysis requires $E_\phi=o(\epsilon^{-1})$, so this residual does not compete with the leading $O(\epsilon^{-1})$ interfacial terms for bounded distributions.
In the bulk-gas limit, $\phi\to1$, $\delta_\epsilon\to0$, and $E_\phi\to0$, and Eq.~\eqref{eqDIKM} reduces to Eq.~\eqref{eqShakhovEq}.

In addition to prescribing the incident and emitted distributions, the DIKM determines the mass, momentum, and energy transferred between the gas and solid through velocity moments of the boundary-flux operator $\calR$.
The local exchange rates per unit interfacial area are
\begin{equation}
    \begin{bmatrix}j_\epsilon\\\bm m_\epsilon\\e_\epsilon\end{bmatrix}
    =\int
    \begin{bmatrix}1\\\bm\xi\\(\xi^2+\eta^2+\zeta^2)/2\end{bmatrix}
    \mathcal  R\diff\bm\varXi.
    \label{eq:interface_moments}
\end{equation}
Here, $j_\epsilon, \bm m_\epsilon$ and $e_\epsilon$ are the interfacial mass, momentum, and energy fluxes, respectively.
Positive values represent transfer to the gas, and multiplication by $\delta_\epsilon$ gives rates per unit computational volume.
In particular,
\begin{equation}
    j_\epsilon=-J_{\rm abs}.
    \label{eq:operator_mass_loss}
\end{equation}
Taking moments of Eq.~\eqref{eqDIKM} gives
\begin{equation}
    \partial_t
    \left[
    \phi
    \begin{pmatrix}
    \rho \\
    \rho\bm u \\
    \rho E
    \end{pmatrix}
    \right]
    +
    \nabla\cdot
    \left[
    \phi
    \begin{pmatrix}
    \rho\bm u \\
    \rho\bm u\bm u+\bm P \\
    \rho E\bm u+\bm P\cdot\bm u+\bm q
    \end{pmatrix}
    \right]
    =
    \delta_\epsilon
    \begin{pmatrix}
    j_\epsilon \\
    \bm m_\epsilon \\
    e_\epsilon
    \end{pmatrix}.
\end{equation}
These balances express the interfacial exchange as a source in the macroscopic transport equations.

The momentum received by the solid in the laboratory frame is $-\bm m_\epsilon$.
At an absorbing moving wall, momentum input includes transport by the absorbed mass, so this contribution requires separation from traction measured relative to the wall.
To separate mechanical traction from momentum carried by absorbed mass, define
\begin{equation}
 \bm t_\epsilon=-\int(\bm\xi-\bm u_\epsilon)\mathcal  R\diff\bm\varXi
 =-\bm m_\epsilon+\bm u_\epsilon j_\epsilon.
 \label{eq:interface_traction}
\end{equation}
The normal load and tangential traction are $-\bm t_\epsilon\cdot\bm n_\epsilon$ and $(\bm I-\bm n_\epsilon\bm n_\epsilon)\cdot\bm t_\epsilon$, respectively.

Energy exchange requires the same frame distinction: laboratory-frame energy transfer includes the contribution associated with wall motion.
The molecular energy delivered to the solid in its local rest frame is
\begin{align}
    q_\epsilon
    &=-\frac12\int\left(|\bm\xi-\bm u_\epsilon|^2+\eta^2+\zeta^2\right)
    \mathcal  R\diff\bm\varXi\notag\\
    &=-e_\epsilon+\bm u_\epsilon\cdot\bm m_\epsilon
    -\frac12|\bm u_\epsilon|^2j_\epsilon.
    \label{eq:wall_heat_flux}
\end{align}
It is positive for energy entering the solid.
At a stationary non-absorbing wall this is the net wall heat flux.
With absorption, it may also include the energy carried by deposited molecules.
Latent heat, binding energy, and solid conduction require an additional material-energy model.
It is therefore distinguished from the gas heat-flux vector $\bm q$ in Eq.~\eqref{eqDF2tau}.

For one solid body, its absorbed mass rate, momentum-input rate, torque about its center $\bm X_b$, and rest-frame energy-input rate are evaluated by
\begin{align}
    \dot M_{\epsilon}&=-\int_\Omega\delta_\epsilon j_\epsilon\diff V,
    \\
    \bm F_\epsilon&=-\int_\Omega\delta_\epsilon\bm m_\epsilon\diff V,
    \label{eq:body_mass_force}
    \\
    \bm T_\epsilon&=-\int_\Omega\delta_\epsilon(\bm x-\bm X_b)\times\bm m_\epsilon\diff V,
    \\
    \dot Q_\epsilon&=\int_\Omega\delta_\epsilon q_\epsilon\diff V.
    \label{eq:body_torque_heat}
\end{align}
The integrals are restricted to that body's layer if several solids are present.
Eq.~\eqref{eq:surface_measure} relates these volume integrals to their sharp-interface counterparts.

\subsection{Matched-asymptotic sharp-interface limit}
\label{subsec:asymptotic}

\begin{figure}[!htp]
    \centering
    \includegraphics[width=0.5\textwidth]{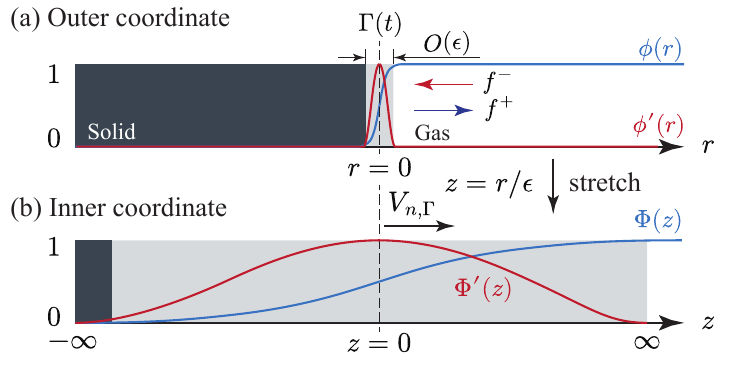}
    \caption{
    Diagram of outer and inner coordinate used in matched asymptotic analysis.
    }
    \label{figAsympAnalysis}
\end{figure}

The purpose of the asymptotic analysis is to determine whether the distributed removal and emission recover the incident and emitted gas-side distributions as the layer thickness vanishes.
For the local analysis across the interface, assume that $\Gamma(t)$ is smooth, its radii of curvature and tangential variation scales are large compared with $\epsilon$, and the wall fields have smooth extensions.
The stretched coordinate $z=r/\epsilon$ resolves the rapid normal variation within the layer, while tangential coordinates $\bm s$ describe variation along the interface.
The coordinates for inner and outer region are sketched in Fig.~\ref{figAsympAnalysis}.

Let $\bm s\in\Gamma(t)$ denote the foot point of the local normal coordinate and write $\bm x=\bm s+\epsilon z\bm n_\Gamma(\bm s,t)$.
We distinguish the inner representations from the physical fields using
\begin{equation}
    \begin{aligned}
        F(z,\bm s,\bm\xi,\bm\eta,\bm\zeta,t;\epsilon)
        &=f(\bm x,\bm\xi,\bm\eta,\bm\zeta,t;\epsilon),\\
        \Phi(z,\bm s,t;\epsilon)&=\phi(\bm x,t;\epsilon).
    \end{aligned}
    \label{eq:inner_representations}
\end{equation}
Both fields are expanded at fixed inner coordinates as
\begin{equation}
    \begin{aligned}
        F&=F_0+\epsilon F_1+O(\epsilon^2),\\
        \Phi&=\Phi_0+\epsilon\Phi_1+O(\epsilon^2).
    \end{aligned}
    \label{eq:inner_expansion}
\end{equation}
$\Phi_0(z,\bm s,t)$ is monotone in $z$, with $\lim_{z\to-\infty}\Phi_0=0$ and $\lim_{z\to+\infty}\Phi_0=1$.
Away from the interface, the corresponding outer expansions are
\begin{equation}
    \begin{aligned}
        f&=f_0+\epsilon f_1+O(\epsilon^2)
        \\
        \phi&=\phi_0+\epsilon\phi_1+O(\epsilon^2)
        .
    \end{aligned}
    \label{eq:outer_expansions}
\end{equation}
In the overlap region, $\epsilon\ll |r|\ll L$, one has $z=r/\epsilon$ large while $r$ remains small compared with a macroscopic variation length $L$.
Expanding the outer coefficients about $r=0$ and comparing equal powers of $\epsilon$ gives the gas-side matching conditions
\begin{equation}
    \begin{aligned}
        \lim_{z\to+\infty}F_0
        &=\lim_{r\to+0}f_0,
    \end{aligned}
    \label{eq:gas_matching_conditions}
\end{equation}
There is no physical gas distribution to prescribe in the solid.
The phase field is defined on both sides, so the same comparison gives, separately for the gas and solid limits,
\begin{equation}
    \begin{aligned}
        \lim_{z\to\pm\infty}\Phi_0
        &=\lim_{r\to\pm0}\phi_0.
    \end{aligned}
\end{equation}
A bounded inner continuation of $F_0$ is required, which gives the solid-side condition
\begin{equation}
    \lim_{z\to-\infty}\Phi_0 F_0=0.
    \label{eq:solid_weighted_matching}
\end{equation}
Let $V_{n,\Gamma}=\bm u_\Gamma \cdot\bm n_\Gamma$ be the normal velocity of the geometric interface.
For a signed-distance coordinate, $\partial_t r=-V_{n,\Gamma}+O(r)$ and $\nabla r=\bm n_\Gamma+O(r)$ near $\Gamma(t)$.
Acting on a bounded smooth inner field while holding the physical position fixed, the chain rule gives the leading derivatives $\partial_t=-V_{n,\Gamma}\epsilon^{-1}\partial_z+O(1)$ and $\nabla=\bm n_\Gamma\epsilon^{-1}\partial_z+O(1)$.
Moreover, $\delta_\epsilon=\Phi_0'/\epsilon+O(1)$ and $\bm n_\epsilon=\bm n_\Gamma+O(\epsilon)$ wherever $\Phi_0'>0$.
The kinematic residual then has the leading form
\begin{equation}
    E_\phi
    =\frac{\Phi_0'}{\epsilon}
    \left(\bm u_\Gamma\cdot\bm n_\Gamma-V_{n,\Gamma}\right)
    +O(1).
    \label{eq:inner_kinematic_residual}
\end{equation}
Thus $E_\phi=o(\epsilon^{-1})$ requires $V_{n,\Gamma}=\bm u_\Gamma\cdot\bm n_\Gamma$ at leading order wherever the profile changes.
The chain rule applied to the weighted distribution now yields
\begin{equation}
    \partial_t(\phi f)+\bm\xi\cdot\nabla(\phi f)
    =\frac{c_n}{\epsilon}\partial_z(\Phi_0 F_0)
    +o(\epsilon^{-1}),
    \qquad
    c_n=(\bm\xi-\bm u_\Gamma)\cdot\bm n_\Gamma.
    \label{eq:app_inner_transport_combined}
\end{equation}
The time derivative supplies $-V_{n,\Gamma}\partial_z(\Phi_0F_0)/\epsilon$, so the transport coefficient is the molecular velocity relative to the moving wall.
Assume bounded inner distributions and wall states, $Q^{\rm S}=O(1)$, and a nonzero leading relative normal velocity $c_n$ for each velocity under consideration.
Substituting the expansions into Eq.~\eqref{eqDIKM} and equating the scale of $\epsilon^{-1}$ gives
\begin{equation}
    c_{n}\partial_z(\Phi_0 F_0)
    =\Phi_0'\mathcal R_0,
    \label{eq:app_inner_leading_balance}
\end{equation}
where $\mathcal R_0$ is the leading term of $\mathcal R$.

For $c_n<0$, the operator in Eq.~\eqref{eq:local_boundary_operator} has $\mathcal R_0=c_nF_0$.
Substitution into Eq.~\eqref{eq:app_inner_leading_balance} and expansion of the product derivative give $c_n(\Phi_0'F_0+\Phi_0\partial_zF_0)=c_n\Phi_0'F_0$, hence $\Phi_0\partial_zF_0=0$.
Wherever $\Phi_0>0$, the incident $F_0$ is independent of $z$.
The gas-side matching condition therefore yields
\begin{equation}
    F_0=f^-,\qquad c_{n}<0.
    \label{eq:app_incident_matching}
\end{equation}
Since the wall fields and $\beta$ are constant along normals at leading order, the local normalization becomes
\begin{equation}
    \lim_{\epsilon\to0}\rho_\epsilon
    =
    \frac{(1-\beta)\displaystyle\int_{c_{n}<0}|c_{n}|f^-\diff\bm\varXi}
    {\displaystyle\int_{c_{n}>0}c_{n}\hat f_\Gamma^{\rm M}\diff\bm\varXi}
    =\rho_\Gamma.
    \label{eq:inner_reflected_density}
\end{equation}
With the incident state fixed by matching, the emitted balance determines whether the wall Maxwellian is recovered without a residual solid-side contribution.
For $c_n>0$, Eqs.~\eqref{eq:local_boundary_operator} and \eqref{eq:inner_reflected_density} give
\begin{equation}
    \mathcal R_0=c_n\rho_\Gamma\hat f_\Gamma^{\rm M}.
\end{equation}
Dividing Eq.~\eqref{eq:app_inner_leading_balance} by $c_n$ and collecting terms gives $\partial_z[\Phi_0(F_0-\rho_\Gamma\hat f_\Gamma^{\rm M})]=0$.
Integration with respect to $z$ therefore yields
\begin{equation}
    \Phi_0 F_0
    =\Phi_0\rho_\Gamma\hat f_\Gamma^{\rm M}+C,
    \label{eq:inner_emitted_integral}
\end{equation}
Taking $z\to-\infty$ in Eq.~\eqref{eq:inner_emitted_integral}, the bounded solid-side continuation gives $\Phi_0F_0\to0$, while $\Phi_0\rho_\Gamma\hat f_\Gamma^{\rm M}\to0$, which gives $C=0$.
Because $\Phi_0>0$ in the transition region, Eq.~\eqref{eq:inner_emitted_integral} now gives $F_0=\rho_\Gamma\hat f_\Gamma^{\rm M}$ there for emitted velocities.
Its gas-side limit, together with Eq.~\eqref{eq:gas_matching_conditions}, recovers
\begin{equation}
    f^+=\rho_\Gamma\hat f_\Gamma^{\rm M},\qquad c_{n}>0.
    \label{eq:app_emitted_matching}
\end{equation}
Together, Eqs.~\eqref{eq:app_incident_matching} and \eqref{eq:app_emitted_matching} recover the moving absorbing diffuse-reflection condition.
In particular, multiplying the incident flux by $\beta$ gives the absorbed flux in Eq.~\eqref{eq:sharp_absorbed_flux}, while the emitted flux is the remaining fraction $1-\beta$ by Eq.~\eqref{eq:inner_reflected_density}.

Curvature affects the next-order tangential geometry and the normal-coordinate volume measure.
In particular,
\begin{equation}
 \diff V=\left[1+\epsilon z\kappa_\Gamma
 +O\!\left((\epsilon z)^2\kappa_*^2\right)\right]
 \epsilon\diff z\diff A,
 \label{eq:curvature_measure}
\end{equation}
where $\kappa_\Gamma=\nabla\cdot\bm n_\Gamma$ and $\kappa_*$ bounds the absolute principal curvatures locally.
Since $\delta_\epsilon=\Phi_0'/\epsilon+\Phi_1'+O(\epsilon)$, multiplying by the normal-coordinate Jacobian shows explicitly how the profile and curvature enter the surface measure.
For a normally constant integrand, and assuming integrable profile derivatives with finite moments so that the expansion may be integrated,
\begin{equation}
    \begin{aligned}
        \int_{-\infty}^{\infty}\delta_\epsilon\,\diff r
        &=1+\epsilon\int_{\mathbb R}\Phi_1'\diff z+O(\epsilon^2),\\
        \int_{-\infty}^{\infty}\delta_\epsilon J\,\diff r
        &=1+\epsilon\int_{\mathbb R}\Phi_1'\diff z\\
        &\quad+\epsilon\kappa_\Gamma\int_{\mathbb R}z\Phi_0'\diff z
        +O(\epsilon^2).
    \end{aligned}
    \label{eq:surface_measure_first_correction}
\end{equation}
Thus the first profile correction is governed by the jump of $\Phi_1$, while the geometric correction is governed by the first moment of $\Phi_0'$.
If the phase field retains an exact unit jump at each $\epsilon$, the integral of $\Phi_1'$ vanishes, but this does not require $\Phi_1$ itself to vanish.
These $O(\epsilon)$ measure terms do not enter the leading $O(\epsilon^{-1})$ inner kinetic balance in Eq.~\eqref{eq:app_inner_leading_balance}.
The symmetric profile in Eq.~\eqref{eqInitPhi} has $\int_{\mathbb R}z\Phi_0'(z)\diff z=0$, so its first curvature contribution cancels for a normally constant integrand.
An asymmetric admissible profile generally has a nonzero first moment.
The leading matching relations identify the boundary law recovered by the inner balance.
The normal constancy of the incident distribution also motivates the reconstruction used in the numerical method.

\subsection{Rigid-body motion and interface evolution}
\label{subsec:body_dynamics}

For a freely moving solid, the interfacial loads determine the motion, and that motion in turn specifies the geometry and wall velocity entering the boundary operator.
For a non-absorbing rigid body of fixed mass $M_b$ and inertia tensor $\bm I_b$ in the laboratory frame, the Newton--Euler equations are
\begin{equation}
    \dot{\bm X_b}=\bm U_b,\qquad
    M_b \dot{\bm U_b}=\bm F_\epsilon+\bm F_{e},\qquad
    \frac{\diff}{\diff t}(\bm I_b\bm\omega_b)=\bm T_\epsilon+\bm T_{e}.
    \label{eq:newton_euler}
\end{equation}
Here $\bm U_b$ and $\bm\omega_b$ are translational and angular velocities, and $\bm F_{e}$ and $\bm T_{e}$ are external force and torque.
For planar motion in the $xy$-plane, let $\theta(t)$ denote the orientation angle of a body-fixed reference axis measured counterclockwise from the laboratory $x$-axis.
With $\bm e_z$ the fixed unit vector normal to the plane, the angular velocity is $\bm\omega_b=\omega_b\bm e_z$, where the signed scalar $\omega_b$ is positive for counterclockwise rotation.
The rotational kinematics and the planar form of the angular-momentum balance in Eq.~\eqref{eq:newton_euler} are
\begin{equation}
    \dot\theta=\omega_b,\qquad
    I_b\dot\omega_b=T_{\epsilon,z}+T_{e,z},
    \label{eq:planar_rotation}
\end{equation}
where $I_b$ is the constant moment of inertia about the axis through $\bm X_b$ parallel to $\bm e_z$, $T_{\epsilon,z}=\bm T_\epsilon\cdot\bm e_z$, and $T_{e,z}=\bm T_e\cdot\bm e_z$.
Given the initial orientation $\theta(0)=\theta_0$, the angle evolves as $\theta(t)=\theta_0+\int_0^t\omega_b(t')\diff t'$.
Together, $\bm X_b(t)$ and $\theta(t)$ specify the position and orientation used to reconstruct the planar rigid-body geometry.
The diffuse wall velocity is
\begin{equation}
    \bm u_\epsilon(\bm x,t)=\bm U_b(t)+\bm\omega_b(t)\times[\bm x-\bm X_b(t)].
    \label{eq:rigid_wall_velocity}
\end{equation}
Rigid-body motion preserves signed distance, so the phase field can be reconstructed from the current position and orientation using Eq.~\eqref{eqInitPhi}, and equivalently, it satisfies $\partial_t\phi+\bm u_\epsilon\cdot\nabla\phi=0$.

\section{Numerical method}
\label{secSolveMethod}

The DIKM developed above is now discretized to simulate gas flows with stationary or moving boundaries.
Operator splitting separates transport step, advanced by DUGKS, from interfacial exchange, treated through a local source update.
The source update preserves non-negativity and quantifies interfacial mass, momentum, and heat transfer through the corresponding discrete exchange quantities.

\subsection{Reduced distributions and velocity discretization}
\label{subsec:weighted_dist_func}

The distribution function $f(\bm x,\bm\xi,\bm\eta,\bm\zeta,t)$ contains unresolved translational and internal coordinates that need not be discretized explicitly for the present spatial configuration.
Integrating over these coordinates gives reduced distributions
\begin{equation}
    g=\int f\diff\bm\eta\diff\bm\zeta,\qquad
    h=\int(\eta^2+\zeta^2)f\diff\bm\eta\diff\bm\zeta
    ,
    \label{eq:weighted_reduced_variables}
\end{equation}
where $g$ retains the mass and resolved momentum information and $h$ retains their contribution to molecular energy.
Letting
\begin{equation}
    \label{eqG_AND_H}
    G=\phi g,\quad H=\phi h,
\end{equation}
the reduced distributions are then weighted by the phase field to incorporate the diffuse solid--gas representation into the evolved kinetic variables.
These weighted variables are used in the transport update instead of original physical distributions, while gas moments and wall incident fluxes are recovered from the underlying reduced distributions in active cells.
The reduced targets obtained from Eq.~\eqref{eqShakovDF} are
\begin{align}
    g^{\rm eq}&=\frac{\rho}{(2\pi RT)^{D/2}}\exp[-c^2/(2RT)],\notag\\
    h^{\rm eq}&=(K+3-D)RTg^{\rm eq},\label{eq:reduced_equilibria}\\
    g^{\rm S}&=g^{\rm eq}\left[1+B\left(\frac{c^2}{RT}-D-2\right)\right],
    \label{eq:reduced_shakhov_g}\\
    h^{\rm S}&=h^{\rm eq}+BRTg^{\rm eq}
    \left[\left(\frac{c^2}{RT}-D\right)(K+3-D)-2K\right],
    \label{eq:reduced_shakhov_h}
\end{align}
where $B=(1-\PrNum)(\bm c\cdot\bm q)/(5pRT)$.
The conservative moments, pressure tensor, and heat flux follow as
\begin{gather}
    \begin{bmatrix}\rho\\\rho\bm u\\\rho E\end{bmatrix}
    =\int\begin{bmatrix}g\\\bm\xi g\\(\xi^2g+h)/2\end{bmatrix}\diff\bm\xi,\notag\\
    \bm P=\int\bm c\bm c g\diff\bm\xi,
    \quad \bm q=\frac12\int\bm c(c^2g+h)\diff\bm\xi.
    \label{eq:reduced_moments}
\end{gather}

Velocity-space discretization is formulated independently of the discretization in physical space and time.
For a polynomial $a(\bm \xi)$, numerical quadrature for kinetic moments uses abscissa $\bm\xi_k$ and coefficients $w_k$:
\begin{equation}
    \int a(\bm\xi)\psi(\bm\xi) \diff\bm\xi
    \simeq
    \sum_{k=1}^{N_\xi} w_k a(\bm\xi_k) \psi(\bm\xi_k),
    \label{eq:velocity_quadrature}
\end{equation}
where $\psi$ stands for $g$ or $h$, and $N_\xi$ is the number of discrete velocities.
To simplify all subsequent velocity sums, absorb the quadrature coefficient into the discrete distributions
\begin{equation}
    g_k =w_k g(\bm\xi_k), 
    \quad h_k= w_k h(\bm\xi_k).
    \label{eq:quadrature_weighted_populations}
\end{equation}

Several quadrature rules can be employed to obtain quadrature nodes and weights \cite{GuoZL2013DUGKS,ChenJF2019}.
Gauss--Hermite (GH) quadrature uses the roots of the Hermite polynomials as nodes for integration.
It requires a Gaussian exponential weight, which matches the form of a Maxwellian, making GH quadrature efficient for near-equilibrium distributions.
Half-range Gauss--Hermite (hGH) quadrature constructs Gaussian rules on $[0,\infty)$ and mirrors them onto $(-\infty,0]$.
Newton--Cotes (NC) composite quadrature uses a truncated velocity box and equally spaced nodes.
On a nonuniform or unstructured velocity mesh, more nodes can be placed near the free-stream and wall distribution centers, with fewer nodes in the tails \cite{ChenJF2019}.
Each quadrature weight equals the volume of the corresponding velocity cell.
Specific quadrature rules are given with the numerical cases.
For notational simplicity, the index $k$ is omitted in the following and is displayed only in explicit velocity sums or where needed to distinguish nodes.

\subsection{Numerical cutoff and operator splitting}
\label{subsec:regularized_split}

Removing the phase field from $G$ and $H$ involves division by $\phi$, whose tanh tails approach zero inside the solid.
To avoid zero-division and taper the source near solid cells, a cutoff $\phi_{\rm cut} = 10^{-3}$ is applied to obtain effective order parameter
\begin{equation}
    \phi_{\rm eff}=\max(\phi,\phi_{\rm cut}).
    \label{eq:effective_phi}
\end{equation}
A cell is termed active where $\phi>\phi_{\rm cut}$ and inactive otherwise, and gas variables are recovered only in active cells.
The numerical source uses $g_{\rm eff}=G/\phi_{\rm eff}$ and $h_{\rm eff}=H/\phi_{\rm eff}$.
Whenever $\phi$ is updated, the regularization in Eq.~\eqref{eq:effective_phi} is applied.
For brevity, the subsequent numerical formulas omit the superscript or subscript ``${\rm eff}$''.

Let $\Psi$ stand for either $G$ or $H$, whose collision targets $\Psi^{\rm S}$ denote $G^{\rm S}=\phi g^{\rm S}$ or $H^{\rm S}=\phi h^{\rm S}$.
The reduced equations are
\begin{equation}
    \dfrac{\partial \Psi}{\partial t}+\bm\xi\cdot\nabla\Psi=Q_\Psi+S_\Psi,
    \qquad Q_\Psi=-\frac{\Psi-\Psi^{\rm S}}{\tau},
    \label{eq:weighted_psi_equation}
\end{equation}
where
\begin{equation}
    \begin{bmatrix}S_G\\S_H\end{bmatrix}
    =\delta_\epsilon\left[
    -\chi^-|c_{n}|
    \begin{pmatrix}G/\phi\\H/\phi\end{pmatrix}
    +\chi^+ c_{n}\rho_\epsilon
    \begin{pmatrix}\hat g_\epsilon^{\rm eq}\\\hat h_\epsilon^{\rm eq}\end{pmatrix}
    \right].
    \label{eq:reduced_sources}
\end{equation}
Here $\hat g_\epsilon^{\rm eq}$ and $\hat h_\epsilon^{\rm eq}$ are the reduced wall Maxwellians at unit density
\begin{gather}
    \hat g_\epsilon^{\rm eq}
    =\frac{1}{(2\pi RT_\epsilon)^{D/2}}
    \exp\!\left[-\frac{|\bm\xi-\bm u_\epsilon|^2}{2RT_\epsilon}\right],
    \\
    \hat h_\epsilon^{\rm eq}
    =(K+3-D)RT_\epsilon\hat g_\epsilon^{\rm eq}.
    \label{eq:discrete_wall_maxwellians}
\end{gather}

Using the Strang-splitting method, Eq.~\eqref{eq:weighted_psi_equation} is split into a symmetric source--transport--source composition
\begin{equation}
    \begin{cases}
        \Psi^{(1)}=\mathcal S_{\Delta t/2}(\Psi^n),\\
        \Psi^{(2)}=\mathcal D_{\Delta t}(\Psi^{(1)}),\\
        \Psi^{n+1}=\mathcal S_{\Delta t/2}(\Psi^{(2)}).
    \end{cases}
    \label{eq:strang_composition}
\end{equation}
The operator $\mathcal S$ advances the interfacial source and $\mathcal D$ advances advection and collision.
The transport substep is computed by the conservative DUGKS method, while the source substep uses a positivity-preserving update with conservative accounting of gas--solid exchange.

\subsection{DUGKS transport update}
\label{subsec:dugks_step}

The operator $\mathcal D_{\Delta t}$ in Eq.~\eqref{eq:strang_composition} advances $\partial_t\Psi+\bm\xi\cdot\nabla\Psi=Q_\Psi$ by the discrete unified gas kinetic scheme (DUGKS) \cite{GuoZL2013DUGKS,GuoZL2015DUGKS}.
Let cell $j$ have measure $V_j$ and center $\bm x_j$, and let $b\subset\partial V_j$ denote a face with center $\bm x_b$, area $A_b$, and outward unit normal $\bm n_{b}$.
The superscripts $(1)$ and $(2)$ identify the input and output of this transport substep.

Integrating over $V_j$ and a time interval $\Delta t$, with midpoint quadrature for transport and trapezoidal quadrature for collision, gives an explicit update
\begin{equation}
    \tilde\Psi_j^{(2)}=\tilde\Psi_j^{+,(1)}
    -\frac{\Delta t}{V_j}\sum_{b}
    A_b(\bm\xi\cdot\bm n_{b})\Psi_b^{n+1/2},
    \label{eq:dugks_main_update}
\end{equation}
in terms of the auxiliary distributions
\begin{equation}
    \tilde\Psi=\Psi-\frac{\Delta t}{2}Q_\Psi,
    \qquad
    \tilde\Psi^+=\Psi+\frac{\Delta t}{2}Q_\Psi.
    \label{eq:tilde_definitions}
\end{equation}
With $Q_\Psi=-(\Psi-\Psi^{\rm S})/\tau$, the transformation relationships are
\begin{align}
    \Psi&=\frac{2\tau}{2\tau+\Delta t}\tilde\Psi
    +\frac{\Delta t}{2\tau+\Delta t}\Psi^{\rm S},
    \label{eq:inverse_tilde_relation}\\
    \tilde\Psi^+&=\frac{2\tau-\Delta t}{2\tau+\Delta t}\tilde\Psi
    +\frac{2\Delta t}{2\tau+\Delta t}\Psi^{\rm S}.
    \label{eq:tilde_postcollision_relation}
\end{align}
The relaxation time and target Shakhov distributions in these relations are evaluated from the local gas macroscopic state and consistent time stage.

To construct $\Psi_b^{n+1/2}$, DUGKS couples transport and collision along a molecular characteristic of duration a half of time step $h_t=\Delta t/2$.
Introduce a second pair of auxiliary distributions,
\begin{equation}
    \bar\Psi=\Psi-\frac{h_t}{2}Q_\Psi,
    \qquad
    \bar\Psi^+=\Psi+\frac{h_t}{2}Q_\Psi,
    \label{eq:bar_definitions}
\end{equation}
Trapezoidal integration of collision along the characteristic lines gives
\begin{equation}
    \bar\Psi(\bm x_b,t_n+h_t)
    =\bar\Psi^+(\bm x_b-h_t\bm\xi,t_n).
    \label{eq:face_distribution_recovery}
\end{equation}
This relation provides the usual DUGKS characteristic construction.
In the diffuse layer its spatial approximation is also required to account for the variation of $\phi$.
Even a spatially uniform underlying gas distribution produces nonuniform $G$ and $H$ across that layer, so directly limiting their slopes would mix gas-state and geometric variations.
Instead, form the effective initial distributions $\bar g_j^+=\bar G_j^{+,(1)}/\phi_j^n$ and $\bar h_j^+=\bar H_j^{+,(1)}/\phi_j^n$ in active cells.
For either effective distribution $\bar\psi^+\in\{\bar g^+,\bar h^+\}$, reconstruct its value at the characteristic foot as
\begin{equation}
    \bar{\psi}^+_b =\bar\psi^+_j+(\bm x_b-h_t\bm\xi-\bm x_j)\cdot\bm\sigma_j,
    \label{eq:effective_reconstruction}
\end{equation}
where $j$ is the upwind cell containing the foot and $\bm\sigma_j$ is the limited gradient of $\bar\psi^+$.
For a Cartesian coordinate direction, the van Leer slope is \cite{vanLeer1977JCP}
\begin{equation}
    \sigma=\begin{cases}
    [\sgn(d_L)+\sgn(d_R)]\dfrac{|d_L||d_R|}{|d_L|+|d_R|},
    &|d_L|+|d_R|>0,\\
    0,&\text{otherwise},
    \end{cases}
    \label{eq:vanleer_slope}
\end{equation}
Here $d_L$ and $d_R$ are the left- and right- sided slopes of $\bar\psi^+$, and $\sgn$ is the sign function.
For an active cell $j$, the gradient component $\sigma_{j,\alpha}$ is set to zero if either neighbor in direction $\alpha$ is inactive ($\phi^n\leq\phi_{\rm cut}$).
Other coordinate directions are limited independently.
Faces shared by two inactive cells carry zero flux.

The reconstructed gas state is converted back to weighted face distributions using the midpoint phase field:
\begin{equation}
    \bar G_b=\phi_b^{n+1/2}\bar g_b,
    \qquad
    \bar H_b=\phi_b^{n+1/2}\bar h_b.
    \label{eq:restore_mask_face}
\end{equation}
The following transformation relations among different distributions will be used in numerical implementation
\begin{gather}
    \bar\Psi^+=\frac{2\tau-h_t}{2\tau+\Delta t}\tilde\Psi
    +\frac{3h_t}{2\tau+\Delta t}\Psi^{\rm S},
    \\
    \Psi=\frac{2\tau \bar\Psi+h_t\Psi^{\rm S}}{2\tau +h_t},
    \label{eq:BarPsi2Psi}
    \\
    \tilde \Psi^+ = \dfrac43 \bar \Psi^+ - \dfrac13 \tilde \Psi.
    \label{eq:BarPlusTilde2TilePlus}
\end{gather}
Recovery of the macroscopic gas state calls for the auxiliary distributions written collectively as $\mathcal A_\Psi=\Psi-aQ_\Psi$, with $a=0$, $\Delta t/2$, or $h_t/2$ corresponding to physical, tilde, or bar distributions, respectively.
In an active cell or at a face admitting gas-state recovery, collision conservation gives
\begin{equation}
    \rho=\frac{\sum_k\mathcal A_{G,k}}{\phi},\qquad
    \rho\bm u=\frac{\sum_k\bm\xi_k\mathcal A_{G,k}}{\phi},\qquad
    \rho E=\frac{\sum_k(\xi_k^2\mathcal A_{G,k}+\mathcal A_{H,k})}{2\phi}.
    \label{eq:unweighted_macros}
\end{equation}
Here $T=(E-|\bm u|^2/2)/c_v$, $c_v=(3+K)R/2$, and $p=\rho RT$.
The relaxation time is then obtained from Eq.~\eqref{eq:viscosity_model}.
Since heat flux and non-equilibrium stress are not collision invariants, their recovery requires the relaxation corrections
\begin{equation}
    \begin{aligned}
    \bm q&=\frac{\sum_k\bm c_k(c_k^2\mathcal A_{G,k}+\mathcal A_{H,k})}
    {2\phi(1+a\PrNum/\tau)},\\
    \bm\Pi&=\frac{\sum_k\bm c_k\bm c_k(\mathcal A_{G,k}-G_k^{\rm eq})}
    {\phi(1+a/\tau)}.
    \end{aligned}
    \label{eq:auxiliary_nonconserved_moments}
\end{equation}
In these expressions, $\bm c_k=\bm\xi_k-\bm u$ and $G_k^{\rm eq}=\phi w_k g^{\rm eq}(\bm\xi_k)$ is the weighted Maxwellian distribution.
The factors $1+a\PrNum/\tau$ and $1+a/\tau$ follow from the Shakhov heat-flux and stress relaxation rates, respectively.
The recovered $\rho$, $\bm u$, $T$, and corrected $\bm q$ determine the targets through Eqs.~\eqref{eq:reduced_equilibria}--\eqref{eq:reduced_shakhov_h}.

The cell reconstruction uses $\phi_j^n$, while a moving interface supplies $\phi_b^{n+1/2}$ through the geometry update in Section~\ref{subsec:discrete_coupling}.
Using $\mathcal A_\Psi=\bar\Psi_b$, $a=h_t/2$, and $\phi=\phi_b^{n+1/2}$ in Eqs.~\eqref{eq:unweighted_macros} and \eqref{eq:auxiliary_nonconserved_moments} gives the face gas state, $\tau_b$, and $\Psi_b^{\rm S}$.
The relation Eq.~\eqref{eq:BarPsi2Psi} then supplies the physical face distribution for Eq.~\eqref{eq:dugks_main_update}.
Inactive faces whose phase field is under cutoff carry zero weighted flux and require no macroscopic recovery.
Thus the order-parameter treatment modifies characteristic reconstruction and face weighting while retaining the DUGKS collision recovery.

After Eq.~\eqref{eq:dugks_main_update}, terminal gas moments are recovered from $\tilde\Psi^{(2)}$ using the updated phase field, and Eq.~\eqref{eq:inverse_tilde_relation} supplies the physical distributions for the second source half-step.
Because the source acts on physical distributions, gas moments, collision targets, and auxiliary distributions are recomputed after each source stage.
On the Cartesian mesh, the time step is chosen as
\begin{equation}
    \Delta t=C_{\rm CFL}
    \max_{k,d} \left[\frac{|\xi_{k,d}| + u_{\max{}}}{\Delta x_{\min,d}}\right]^{-1},
    \label{eq:time_step}
\end{equation}
where $C_{\rm CFL}\in(0,1)$ is the Courant number, $\xi_{k,d}$ is the $d$th component of $\bm\xi_k$, and $\Delta x_{\min,d}$ is the finest cell spacing in direction $d$.
The maximum velocity magnitude $u_{\max}$ is obtained in gas bulk where $\phi > 0.95$.

\subsection{Positivity-preserving source update}
\label{subsec:discrete_source}

The distribution removal treatment in the boundary-flux operator can deplete an incident distribution within a single transport time step.
A direct explicit subtraction may produce negative distributions and lead to numerical instability.
The present source update avoids this restriction by integrating the decay of the incident population exactly and returning the reflected fraction of the mass actually removed.
In each active cell, incident distributions obey
\begin{equation}
    G^{\rm post}=e^{-h_t\lambda}G^{\rm pre},\quad
    H^{\rm post}=e^{-h_t\lambda}H^{\rm pre},
    \quad c_{n}<0,
    \label{eq:positive_incident_removal}
\end{equation}
during a source substep of duration $h_t$, where $\lambda= \delta_\epsilon|c_{n}| / \phi$.
The superscripts ``${\rm pre}$'' and ``${\rm post}$'' denote the states before and after the source substep.
The reflected and absorbed mass per unit cell volume are
\begin{align}
    R_{\rm refl}&=(1-\beta)\sum_{c_{n,k}<0} (G_k^{\rm pre}-G_k^{\rm post}),
    \\
    R_{\rm abs}&=\beta\sum_{c_{n,k}<0} (G_k^{\rm pre}-G_k^{\rm post}).
    \label{eq:total_reflected_mass}
\end{align}
The emitted distributions then follow from
\begin{equation}
    G^{\rm post}=G^{\rm pre}
    +\frac{c_{n}\hat g_{\epsilon}^{\rm eq}}{A_\epsilon}R_{\rm refl},
    \quad
    H^{\rm post}=H^{\rm pre}
    +\frac{c_{n}\hat h_{\epsilon}^{\rm eq}}{A_\epsilon}R_{\rm refl},
    \quad c_{n}>0,
    \label{eq:positive_emitted_supply}
\end{equation}
where
\begin{equation}
    A_\epsilon
    = \sum_{\,c_{n,k}>0}
    c_{n,k}\hat g_{\epsilon,k}^{\rm eq},
    \qquad
    c_{n,k}=(\bm\xi_k-\bm u_\epsilon)\cdot\bm n_\epsilon,
    \label{eq:discrete_wall_flux_normalization}
\end{equation}
For nonnegative input distributions, these updates preserve non-negativity for arbitrary $h_t\lambda$ and satisfy $\sum_k(G_k^{\rm post}-G_k^{\rm pre})=-R_{\rm abs}$.

\subsection{Discrete exchange and moving-body coupling}
\label{subsec:discrete_coupling}

The discrete solid loads needs match the exchange actually applied to the gas during each source stage.
These loads are evaluated by accumulating the distribution changes.
Interfacial exchange uses the increments of the physical distributions, $\Delta G_{j,k}=G_{j,k}^{\rm post}-G_{j,k}^{\rm pre}$ and $\Delta H_{j,k}=H_{j,k}^{\rm post}-H_{j,k}^{\rm pre}$.
The solid impulse and energy input in a source substep are
\begin{align}
    \Delta\bm P_\epsilon&=-\sum_jV_j\sum_k\bm\xi_k\Delta G_{j,k},
    \label{eq:discrete_impulse}\\
    \Delta\bm L_\epsilon&=-\sum_jV_j(\bm x_j-\bm X_b)\times
    \sum_k\bm\xi_k\Delta G_{j,k},
    \label{eq:discrete_angular_impulse}\\
    \Delta Q_\epsilon&=-\frac12\sum_jV_j\sum_k
    \left(|\bm\xi_k-\bm u_{\epsilon,j}|^2\Delta G_{j,k}+\Delta H_{j,k}\right).
    \label{eq:discrete_heat}
\end{align}
The absorbed mass is $\Delta M_{\epsilon}=\sum_jV_jR_{{\rm abs},j}$.
Summing the two half-step impulses and dividing by $\Delta t$ gives the step-averaged force.

For planar fixed-mass rigid-body motion without external loads, the staggered impulse--drift update is
\begin{align}
    \bm U_b^{n+1/2}
    &= \bm U_b^n + \frac{\Delta \bm P_\epsilon^{(1)}}{M_b},
    &
    \omega_b^{n+1/2}
    &= \omega_b^n + \frac{\Delta L_\epsilon^{(1)}}{I_b},
    \label{eq:first_body_kick}
    \\[3pt]
    \bm X_b^{n+1/2}
    &= \bm X_b^n + \dfrac{\Delta t}{2}\,\bm U_b^{n},
    &
    \theta_b^{n+1/2}
    &= \theta_b^n + \dfrac{\Delta t}{2}\,\omega_b^{n},
    \label{eq:first_body_drift}
    \\[3pt]
    \bm U_b^{n+1}
    &= \bm U_b^{n+1/2}
     + \frac{\Delta \bm P_\epsilon^{(2)}}{M_b},
    &
    \omega_b^{n+1}
    &= \omega_b^{n+1/2}
     + \frac{\Delta L_\epsilon^{(2)}}{I_b}.
    \label{eq:second_body_kick}
    \\[3pt]
    \bm X_b^{n+1}
    &= \bm X_b^n + \Delta t\,\bm U_b^{n+1/2},
    &
    \theta_b^{n+1}
    &= \theta_b^n + \Delta t\,\omega_b^{n+1/2},
    \label{eq:second_body_drift}
\end{align}
After the first impulse, a half drift supplies $\phi^{n+1/2}$ for the transport face flux, while the cell reconstruction uses the initial phase field $\phi^n$.
The remaining half drift supplies $\phi^{n+1}$ for terminal macroscopic recovery.
The second source stage uses this new phase field and the intermediate rigid-body velocity.
External forces and torques can be included through the impulses.
This update conserves source momentum between gas and solid.

Updating the phase field may cause active cells to become inactive, or inactive cells to become active.
Deactivation occurs when $\phi_j^{n+1}\leq\phi_{\rm cut}<\phi_j^n$.
After transport and the full rigid-body drift, auxiliary distributions $\tilde\Psi$ left in inactive cells are redistributed to active neighbors, before conversion to physical distributions.
Let $\mathcal N_j$ denote the set of eligible face-adjacent active recipient cells of an inactive donor cell $j$.
The transfer is implemented as
\begin{equation}
    \alpha_{ij}=\frac{\phi_i}{\sum_{m\in\mathcal N_j}\phi_m},\qquad
    \tilde\Psi_i
    \leftarrow\tilde\Psi_i
    +\sum_{j:\,i\in\mathcal N_j}\alpha_{ij}\tilde\Psi_j,
    \qquad
    \tilde\Psi_j \leftarrow 0,
    \label{eq:residual_transfer}
\end{equation}
Here, $j$ denotes an inactive donor cell and $i \in \mathcal N_j$ an active recipient.
The donors are cleared when its residual is fully assigned to eligible recipients.
The same nonnegative transfer weights are applied to every discrete velocity and to both distribution functions.
On the other hand, a newly active cell satisfies $\phi_i^n\leq\phi_{\rm cut}<\phi_i^{n+1}$.
The transport step first updates the weighted distributions in this cell through any nonzero fluxes across faces shared with active neighboring cells.
Once the updated geometry identifies it as active, the cell receives any eligible residual transfers through Eq.~\eqref{eq:residual_transfer}.
Physical distributions and gas moments are then recovered using $\phi_i^{n+1}$, followed by the second source half-step.

\subsection{Adaptive mesh refinement}
\label{subsec:amr}

Resolving the thin diffuse-interface region on a uniformly fine mesh would incur unnecessary computational cost away from the interface.
Adaptive mesh refinement (AMR) is therefore employed to improve computational efficiency by concentrating fine cells in the diffuse-interface region while retaining coarser cells elsewhere.
The present simulations use block-structured AMR based on the AMReX framework \cite{ZhangWQ2021}.
The mesh consists of nested levels $\ell=0,\ldots,L_{M} - 1$, with rectangular fine-grid patches covering selected coarse-grid regions.
Coarse-level cells can be tagged for refinement based on user-defined criteria, such as the interface location, the values or gradients of flow variables, and subsequently refined according to a prescribed refinement ratio.

All refinement levels advance with a common time step determined by Eq.~\eqref{eq:time_step}.
To provide the reconstruction stencils at patch boundaries, each patch is extended with auxiliary cells, referred to as ghost cells, whose values are supplied by neighboring patches or interpolated from the next coarser level.
After advancement, fine-cell data are averaged onto the coarse cells they cover, and conservation across coarse--fine interfaces requires matching integrated transport fluxes through a shared flux or a flux-register correction \cite{Boyd2026JCP}.

\subsection{Overall algorithm}
\label{subsec:algorithm}

The complete time step combines the two interfacial source subseps with DUGKS transport, optional phase-field evolution, and mesh refinement.
Its stage ordering ensures that reconstruction uses the time-consistent phase field and that solid loads are accumulated from the corresponding source increments.
The overall algorithm is generalized as following
\begin{enumerate}
    \item Initialize the phase field $\phi$, gas macroscopic state $\rho, \rho \bm u$ and $\rho E$, distribution functions $g, h$, discrete velocity set $\{\bm \xi_k\}_{k=1}^{N_\xi}$, and mesh hierarchy $\{\ell\}_0^{L_M-1}$.
    Construct initial $\tilde \Psi$ by Eqs.~\eqref{eqG_AND_H} and \eqref{eq:tilde_definitions}.
    \item Regrid if needed, and compute $\Delta t$ by Eq.~\eqref{eq:time_step}.
    Construct the interface normal $\bm n_\epsilon$ and surface delta $\delta_\epsilon$ by Eq.~\eqref{eq:diffuse_normal} and \eqref{eq:diffuse_geometry}, respectively.
    \item Recover the physical distributions $\Psi$ by Eq.~\eqref{eq:inverse_tilde_relation}.
    Apply the first source half-step by Eqs.~\eqref{eq:positive_incident_removal} and \eqref{eq:positive_emitted_supply}, and accumulate its mass, momentum, angular-momentum, and energy increments by Eqs.~\eqref{eq:discrete_impulse}--\eqref{eq:discrete_heat}.
    \item For a free body, apply the first impulse and drift to the midpoint using Eqs.~\eqref{eq:first_body_kick} and \eqref{eq:first_body_drift}.
    Construct the midpoint phase field $\phi^{n+1/2}$.
    Compute the gas macroscopic states, and form the DUGKS auxiliary distributions $\tilde \Psi$ and $\bar \Psi^+$.
    \item Calculate $\bar \Psi_b^{n+1/2}$ by Eqs.~\eqref{eq:face_distribution_recovery}, \eqref{eq:restore_mask_face}, and \eqref{eq:effective_reconstruction}, and then compute $\Psi_b^{n+1/2}$ by Eq.~\eqref{eq:BarPsi2Psi}.
    Advance the transport substep by Eq.~\eqref{eq:dugks_main_update}.
    \item Complete the body drift or prescribed geometry update using Eq.~\eqref{eq:second_body_drift}.
    Recover physical distributions $\Psi$, and handle cells crossing the activity threshold.
    \item Apply the second source half-step Eq.~\eqref{eq:second_body_kick} and accumulate its exchange increments.
    Apply the second rigid-body impulse and recover the final gas fields and auxiliary distributions by Eqs.~\eqref{eq:discrete_impulse}--\eqref{eq:discrete_heat}.
    \item Synchronize the level data and account for coarse--fine flux balance.
    \item Repeat steps 2 to 8 until reaching steady state or meeting certain exiting conditions.
\end{enumerate}

\section{Results and discussion}
\label{secResult}

The numerical tests examine distinct aspects of the present DIKM:
continuum to rarefied gas flows with momentum and energy transport,
arbitrary molecular absorption probabilities,
planar or curved interfaces,
and stationary or moving free bodies.
Comparisons with reference benchmarks and sharp-interface (SI) calculations assess the model's capability, accuracy, and convergence, as well as the additional effects of the diffuse-interface (DI) treatment.
The final physical vapor deposition (PVD) example extends the simulation to an evolving absorbing surface and demonstrates the DIKM's capability for practical applications in material processes.

\subsection{Planar Couette flow}
\label{subsec:couette}

\begin{figure}[!htp]
    \centering
    \safeincludegraphics[width=0.70\textwidth]{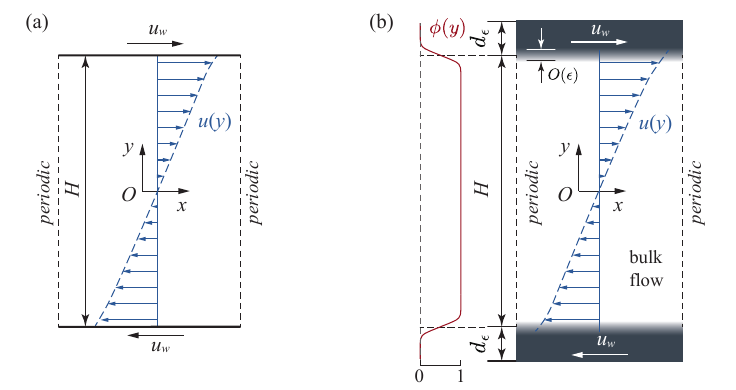}
    \caption{Schematic diagram of planar Couette flow with (a) SI and (b) DI representation.}
    \label{fig:couette_diagram}
\end{figure}

Planar Couette flow serves as a simple but effective verification of momentum transfer between a gas and planar moving walls, without geometric curvature or external forces.
It therefore provides a baseline test of velocity slip and wall shear stress before the diffuse-interface treatment is applied to more complex geometries.
Two plates at $y=\pm H/2$ move in opposite directions with speed $u_w$, as illustrated in Fig.~\ref{fig:couette_diagram}.
The gas is initially stationary with reference density $\rho_0$, and both plates and the initial gas have temperature $T_0$.
Periodic streamwise boundaries are set to make the resolved flow essentially one-dimensional in the wall-normal direction.

The molecular mean free path and Knudsen number are redefined by
\begin{equation}
    \label{eq:couette_mean_free_path}
    \lambda
    =
    \frac{\mu}{\rho}
    \sqrt{\frac{\pi}{2RT}},
\end{equation}
and $\KnNum=\lambda/H$, respectively.
The DI domain includes a solid extension of depth $d_\epsilon$ beyond each plate, and its constant  order parameter is
\begin{equation}
    \label{eqPhiInitCouetteFlow}
    \phi(y) = \dfrac12
    \left[
        \tanh \dfrac{3(2y + H)}{2\epsilon}
        + \tanh \dfrac{3(H-2y)}{2\epsilon}
    \right].
\end{equation}
The wall velocity is prescribed using the diffuse normal $\bm n_\epsilon = (n_{\epsilon,x}, n_{\epsilon,y})$ pointing from solid to gas:
\begin{equation}
    \bm u_\epsilon
    =
    \begin{cases}
        (u_w,0), & n_{\epsilon,y} < 0,\\
        (-u_w,0), & n_{\epsilon,y} > 0.
    \end{cases}
    \label{eq:couette_solid_velocity}
\end{equation}
Thus, the upper and lower interfacial layers impose the velocities of their respective plates while the physical gas gap remains $H$.

Unless otherwise stated, the simulation parameters are $H=1$, $d_\epsilon=0.1$, $u_w=0.1$, $\rho_0=1$, $C_{\rm CFL}=0.5$, and $\epsilon=2\Delta x$.
Knudsen numbers ranging from $10^{-3}$ to $10$ are tested.
At $\KnNum=10^{-3}$, a uniform $2\times256$ mesh and a tensor-product $8\times8$-point half-range Gauss--Hermite (hGH) quadrature are used.
The other stated cases use a $2\times128$ mesh and a $24\times24$-point hGH quadrature.
The discretization parameters were selected based on an independence test, as will be shown later.
Steady state is assessed using $L_2$-norm
\begin{equation}
    E_c(n)
    =
    \frac{\|u_j^{n+1}-u_j^n\|_2}{\|u_j^n\|_2}
    \le10^{-7}.
    \label{eq:convergence_criterion}
\end{equation}
where $n$ denotes the time level and $j$ the sampling cells of $\phi > 0.5$.

\begin{figure}[!htp]
    \centering
    \safeincludegraphics[width=0.8\textwidth]{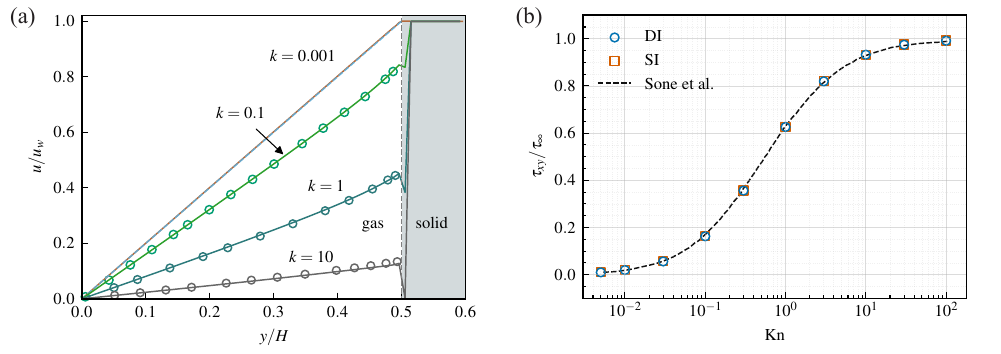}
    \caption{
    Simulation results for planar Couette flow.
    (a) Velocity profiles at different rarefied parameter $k = (\sqrt{\pi}/2) \KnNum$.
    Solid lines are present results while markers denote benchmark solutions from Ref.~\cite{Sone1990}.
    (b) Normalized shear stress $\tau_{xy}/\tau_\infty$ for SI and DI calculation.
    The dashed line denotes the benchmark solution from Ref.~\cite{Sone1990}.
    }
    \label{fig:couette_velocity_stress}
\end{figure}

Fig.~\ref{fig:couette_velocity_stress}(a) plots $u/u_w$ against $y/H$ in the upper half of the domain, with rarefaction parameter $k = (\sqrt{\pi}/2) \KnNum$.
In continuum regime $k=10^{-3}$, the gas velocity is approximately linear and approaches the no-slip profile.
As rarefaction increases, the gas-side velocity at the wall decreases relative to $u_w$, and the departure from the continuum profile becomes more pronounced.
The velocity curves reproduce the same trend as reference markers from Ref.~\cite{Sone1990}.
The rapid transition to the prescribed wall velocity occurs within the diffuse-interface layer and the solid extension.
Localized velocity oscillations occur near the diffuse interface.
These may arise from discretization and splitting errors in the coupled weighted transport and interfacial source updates, particularly when the transition layer spans few cells.
Recovery of velocity from small order-parameter-weighted density and momentum can amplify inconsistencies between these moments.
The good agreement of the bulk velocity profiles and wall shear stresses with the benchmarks suggests that the localized oscillations have limited effects on bulk gas transport.

An assessment of momentum transfer is provided by the wall shear stress.
The non-equilibrium stress introduced in Eq.~\eqref{eqDF2tau} is recovered from
\begin{equation}
    \bm\Pi
    =
    \int
    \bm c\bm c
    \left(f-f^{\rm eq}\right)
    \diff\bm\xi,
    \label{eq:stress_tensor}
\end{equation}
whose wall-parallel component supplies the shear stress $\tau_{xy}$.
For the DI calculation, the wall load is obtained by accumulating the interfacial momentum exchange.
Fig.~\ref{fig:couette_velocity_stress}(b) shows $\tau_{xy}/\tau_\infty$, where $\tau_\infty=-\rho u_w\sqrt{2RT/\pi}$.
The DI and SI results capture the increase toward the free-molecular stress scale and agree well with the reference curve from continuum to free-molecular regime.

\begin{figure}[!htp]
    \centering
    \safeincludegraphics[width=0.8\textwidth]{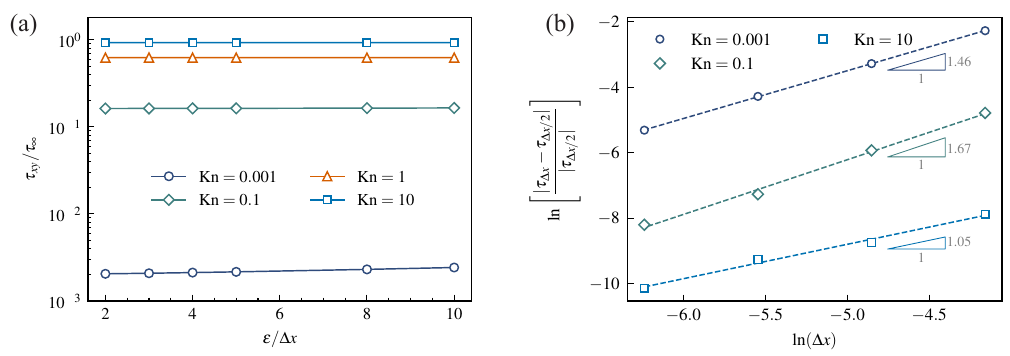}
    \caption{Couette-flow shear-stress sensitivity of interface width and grid size:
    (a) normalized shear stress versus $\epsilon/\Delta x$ with fixed $\Delta x$, and
    (b) convergence rate against space step with fixed $\epsilon/\Delta x$.
    }
    \label{fig:couette_independence}
\end{figure}

Fig.~\ref{fig:couette_independence}(a) shows the dependence of normalized shear stress on $\epsilon/\Delta x$ to guide the choice of interface width.
The shear stress varies little with interface width in the more rarefied cases, whereas the smallest Knudsen number exhibits greater sensitivity at larger $\epsilon/\Delta x$.
These results suggest using an interface width in the range $2\Delta x \leq \epsilon \leq 5\Delta x$.
Fig.~\ref{fig:couette_independence}(b) compares shear stresses on successively refined grids at fixed $\epsilon/\Delta x$ to assess its convergence rate.
The ordinate corresponds to the relative inter-grid difference $E_\tau(\Delta x)=|\tau_{\Delta x}-\tau_{\Delta x/2}|/|\tau_{\Delta x/2}|$.
For two successive spacings, the associated rate can be evaluated as $p_\tau=\ln[E_\tau(\Delta x)/E_\tau(\Delta x/2)]/\ln 2$.
The slopes annotated in the figure are $1.46$, $1.67$, and $1.05$ by linear fit for $\KnNum=0.001$, $0.1$, and $10$, respectively.
Taken together, the results demonstrate the ability of the DIKM to reproduce the principal features of planar shear flow over the tested rarefaction range, and provide evidence of reducing errors as space step $\Delta x$ decreases.

\subsection{Planar Fourier flow}

Planar Fourier flow tests thermal transport and gas--wall energy exchange.
The geometry is the same as in Fig.~\ref{fig:couette_diagram}, but the walls are stationary and maintained at different temperatures.
The present settings assign $T_h=1.25 T_0$ to the upper wall and $T_l=0.75 T_0$ to the lower wall, with the initially uniform gas temperature and reference temperature $T_0$.
The hard-sphere viscosity law with $\omega=0.5$ is used, with Prandtl number $\PrNum = 2/3$.
A uniform $2\times128$ mesh and a $32\times32$-point hGH quadrature are employed at $\KnNum=0.01$, $0.1$, $1$ and $10$.
Other parameters take the same value as those in the Couette flow case.
The convergence criterion monitors temperature variation:
\begin{equation}
    E_c(n)=\frac{\|T_{y,j}^{n+1}-T_{y,j}^{n}\|_2}{\|T_{y,j}^{n}\|_2}
    \le 10^{-7}.
\end{equation}

\begin{figure}[!htp]
    \centering
    \safeincludegraphics[width=0.8\textwidth]{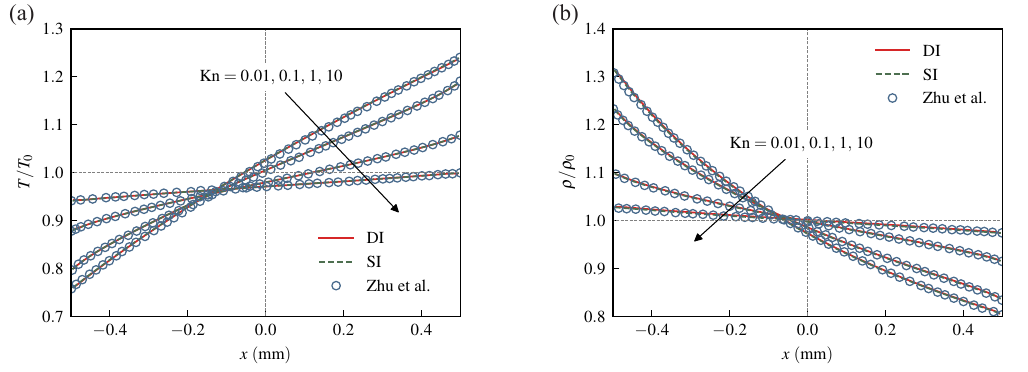}
    \caption{Distribution of (a) temperature and (b) density in steady Fourier flow at different Knudsen numbers $\KnNum = 0.01, 0.1$, $1$ and $10$.
    Circular markers are benchmark solutions from \cite{ZhuLH2021JCP}, dashed lines and  black dot lines are DI and SI results, respectively.
    }
    \label{fig:fourierflow}
\end{figure}

The normalized temperature and density profiles are shown in Fig.~\ref{fig:fourierflow}.
The temperature profiles become flatter as the plotted Knudsen number increases, while the density variation across the gap decreases.
The comparatively larger temperature and density gradients at lower rarefaction are also reproduced.
Temperature and density vary in opposite directions, and the DI and SI profiles follow the same overall trends and closely follow the reference markers \cite{ZhuLH2021JCP}.

\subsection{Semi-infinite absorbing boundary}
\label{subsec:absorbing_boundary}

\begin{figure}[!htp]
    \centering
    \includegraphics[width=0.5\textwidth]{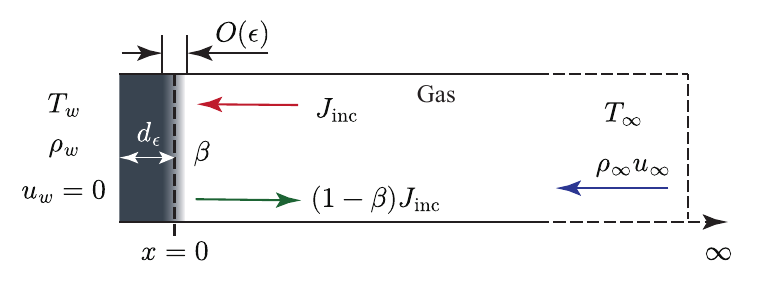}
    \caption{Schematic diagram for the semi-infinite absorbing boundary problem.}
    \label{figAbsorbingBCDiagram}
\end{figure}

This test directly examines whether the diffuse-interface source can realize a prescribed absorption probability and balances the incident, absorbed and reflected fluxes.
As shown in Fig.~\ref{figAbsorbingBCDiagram}, the physical domain is $x\in[0,\infty)$, with a stationary isothermal wall at $x=0$ and far-field quantities $\rho_\infty$, $u_\infty$, and $T_\infty$.
The wall temperature is $T_w=T_\infty$, and absorption draws gas toward the wall, so in steady state $u_\infty<0$ for nonzero absorption probability.
For molecules reflected from the wall, the distribution is
\begin{equation}
    \label{eq:absorbing_wall_distribution}
    f(x=0,\bm\xi,t)
    =
    \frac{\rho_w}{(2\pi RT_w)^{3/2}}
    \exp\left[
        -\frac{\bm\xi^2}{2RT_w}
    \right],
    \qquad
    \xi_x>0,
\end{equation}
The wall density $\rho_w$ is determined by 
\begin{equation}
    \label{eq:absorbing_wall_flux_balance}
    \rho_w\sqrt{\frac{RT_w}{2\pi}}
    =
    -(1-\beta)J_{\rm inc},
    \qquad
\end{equation}
Here the signed incident flux is
\begin{equation}
    J_{\rm inc}=\int_{\xi_x<0}\xi_x f(0,\bm\xi,t)\diff\bm\xi<0.
\end{equation}
The same wall-directed quantity is denoted $J_{\rm out}$ in Ref.~\cite{Frezzotti2012}, where it is viewed as leaving the gas and entering the wall; it is written $J_{\rm inc}$ here, consistently with the terminology above.
At the remote boundary, the populations entering the domain are taken as an equilibrium distribution
\begin{equation}
    f(x=\infty,\bm\xi,t)
    =
    \frac{\rho_\infty}{(2\pi RT_\infty)^{3/2}}
    \exp\left[
        -\frac{|\bm\xi-\bm u_\infty|^2}{2RT_\infty}
    \right],
    \qquad
    \xi_x<0 .
    \label{eq:absorbing_farfield_distribution}
\end{equation}

\begin{figure}[!htp]
    \centering
    \safeincludegraphics[width=0.8\textwidth]{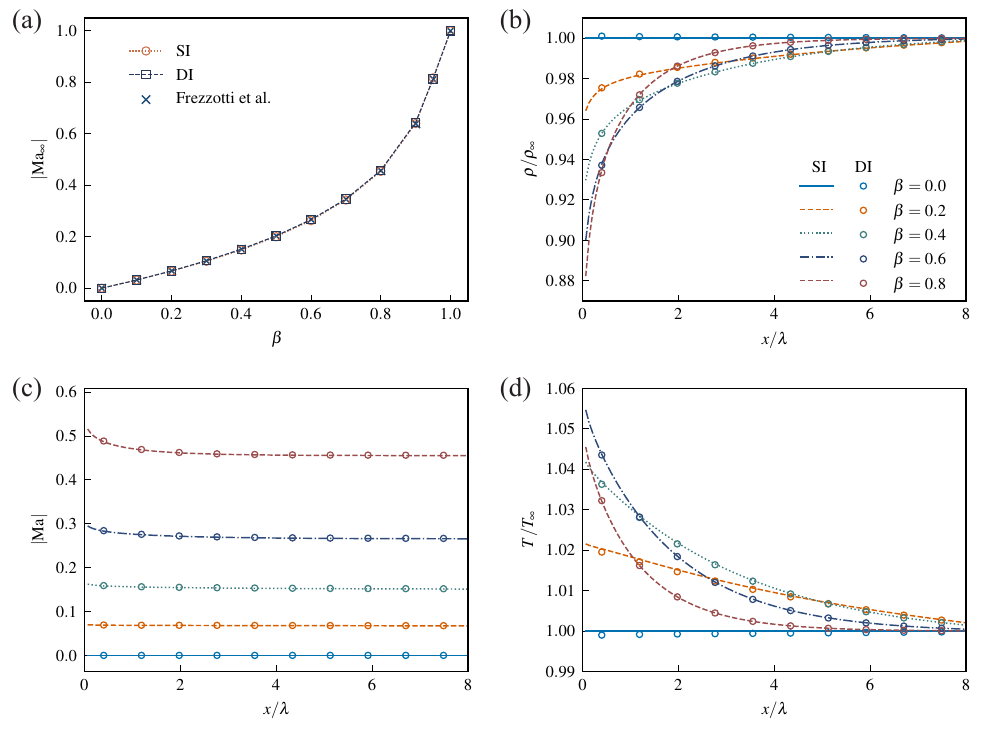}
    \caption{
    (a) Far-field Mach-number magnitude versus prescribed absorption probability, comparing DI,SI, and reference results from Ref.~\cite{Frezzotti2012}.
    Distributions of (b) Density (c) Mach number and (d) temperature in the Knudsen layer for different absorption probabilities.}
    \label{figAbsorbingAlpha}
\end{figure}

The finite computational gas interval is set to $[0,L]$, extended to $[-d_\epsilon,L]$ in the DI calculation with $d_\epsilon=L/12$.
The present settings use $L=4$ for Fig.~\ref{figAbsorbingAlpha}(a) and $L=0.2$ for Fig.~\ref{figAbsorbingAlpha}(b)--(d).
The reference mean free path is $\lambda=\tau\sqrt{\pi RT_w/2}$, and the Prandtl number is $\PrNum=1$.
The calculation uses $128$ cells, a $24\times24$-point hGH quadrature, and $C_{\rm CFL}=0.8$.
The interface width is $\epsilon = 5\Delta x$.
In the absence of shear flow, the far-field Mach number is determined by the absorbing probability and temperature ratio through \cite{Frezzotti2012}
\begin{equation}
    \MaNum_\infty=\MaNum_\infty(\beta,T_\infty/T_w).
\end{equation}
The order parameter is initialized as
\begin{equation}
    \phi(x)=\frac12\left[1+\tanh\left(\frac{3x}{\epsilon}\right)\right].
\end{equation}

Fig.~\ref{figAbsorbingAlpha}(a) shows that the magnitude of the far-field Mach number increases non-linearly with absorption and that the DI and SI values follow close to the reference trend \cite{Frezzotti2012}.
Fig.~\ref{figAbsorbingAlpha}(b)--(d) display normalized density, Mach-number magnitude, and temperature from $\beta=0$ to $0.8$.
The absorbing cases exhibit a density deficit and increased speed near the wall, followed by relaxation toward their far-field values.
Small temperature deviations are localized near the wall, but the temperature profiles follow the same trend as the SI results and rapidly converge towards them.

The balance of fluxes provide a more direct verification of the imposed boundary condition.
At steady state, $J_\infty=\rho_\infty u_\infty=\beta J_{\rm inc}$, and hence $\beta_m=|J_\infty/J_{\rm inc}|$.
The relative error is defined as $\eta_r=|\beta_m-\beta|/\beta$ and listed in Tab.~\ref{tabAbsorbBCFluxCompare}, where $\beta_m$ denotes the measured value of absorption probability.
For $\beta=0.2$, the error is $0.15\%$, while the intermediate rows from $0.4$ to $0.8$ agree to the displayed precision.
The fully absorbing case gives $\beta_m=1.0088$ and a relative discrepancy of $0.88\%$.
Therefore, these data demonstrate correct recovery of the prescribed nonzero absorption probabilities.

\begin{table}[!htp]
    \centering
    \caption{Numerically calculated fluxes and measured absorbing rate of different prescribed absorbing probability $\beta$.}
    \label{tabAbsorbBCFluxCompare}
    \begin{tabular}{ccccc}
        \toprule
        prescribed & incident flux & far-field flux & measured & relative error \\
        $\beta$ & $|J_{\rm inc}|$ & $|J_\infty|$ & $\beta_m = |J_\infty / J_{\rm inc}|$ & $\eta_r (\%)$ \\
        \midrule
        $0.0$ & $0.811554$ & $0.000004$ & $4.93\times10^{-6}$ & ----- \\
        $0.2$ & $0.307632$ & $0.061428$ & $0.1997$ & $0.15$ \\
        $0.4$ & $0.345366$ & $0.138142$ & $0.4000$ & $0.00$ \\
        $0.6$ & $0.405097$ & $0.243058$ & $0.6000$ & $0.00$ \\
        $0.8$ & $0.522004$ & $0.417604$ & $0.8000$ & $0.00$ \\
        $1.0$ & $0.912541$ & $0.920573$ & $1.0088$ & $0.88$ \\
        \bottomrule
    \end{tabular}
\end{table}


The case is further extended to simulate one-dimensional subsonic condensation problems, to compare with the benchmarks in Ref.~\cite{Frezzotti2012}.
In this formulation, the emitted wall Maxwellian is specified by a saturated vapor pressure $p_w=\rho_wRT_w$, rather than by prescribing an independent reflection fraction.
The steady mass balance is
\begin{equation}
    \label{eq:condensation_flux_balance}
    \rho_w\sqrt{\frac{RT_w}{2\pi}}+J_{\rm inc}
    =
    \rho_\infty u_\infty .
\end{equation}
Combining this relation with Eq.~\eqref{eq:absorbing_wall_flux_balance} yields
\begin{equation}\label{eq:absorption_condensation_relation}
    \frac{p_w}{p_\infty}
    =
    \frac{1-\beta}{\beta}
    \sqrt{2\pi\gamma}
    \sqrt{\frac{T_w}{T_\infty}}
    |\MaNum_\infty| .
\end{equation}
Therefore, for $T_w=T_\infty$ and $\beta \in [0,1)$, the far-field velocity is
\begin{equation}
    \label{eq:farfield_velocity_absorption}
    u_\infty
    =
    -\dfrac{\beta}{1-\beta}
        \dfrac{n_w}{n_\infty}
        \sqrt{\dfrac{RT_w}{2\pi}},
\end{equation}
where $n_w$ and $n_\infty$ are the number densities of the same molecular species.

For this comparison, the coordinate is normalized by $\tilde\lambda=\tau\sqrt{8RT_w/\pi}$.
The six far-field parameter sets are listed in Tab.~\ref{tab:subsonic_condensation_cases}, where the sound speed is $c_{s,\infty} = \sqrt{\gamma RT_\infty}$.
A $24\times24$-point hGH quadrature is used except in case (d), which uses $40\times40$ points.
All other simulation parameters are identical to those specified previously.

\begin{table}[!htp]
    \centering
    \small
    \caption{Parameters for the subsonic condensation steady-state cases used for comparison with the reference solutions of Aoki et al.~\cite{Aoki1990}.}
    \label{tab:subsonic_condensation_cases}
    \setlength{\tabcolsep}{30pt}
    \renewcommand{\arraystretch}{1.1}
    \begin{tabular}{cccc}
        \toprule
        Case & $|\MaNum_{\infty}|$ & $p_{\infty}/p_w$ & $T_{\infty}/T_w$ \\
        \midrule
        (a) & 0.50757  & 3.1858 & 0.51220 \\
        (b) & 0.51829  & 3.1356 & 1.0213  \\
        (c) & 0.53276  & 3.2785 & 2.0194  \\
        (d) & 0.50012  & 3.2129 & 4.1124  \\
        (e) & 0.095447 & 1.2088 & 1.0030  \\
        (f) & 0.91185  & 9.8513 & 0.99389 \\
        \bottomrule
    \end{tabular}
\end{table}

\begin{figure}[!htp]
    \centering
    \includegraphics[width=0.95\textwidth]{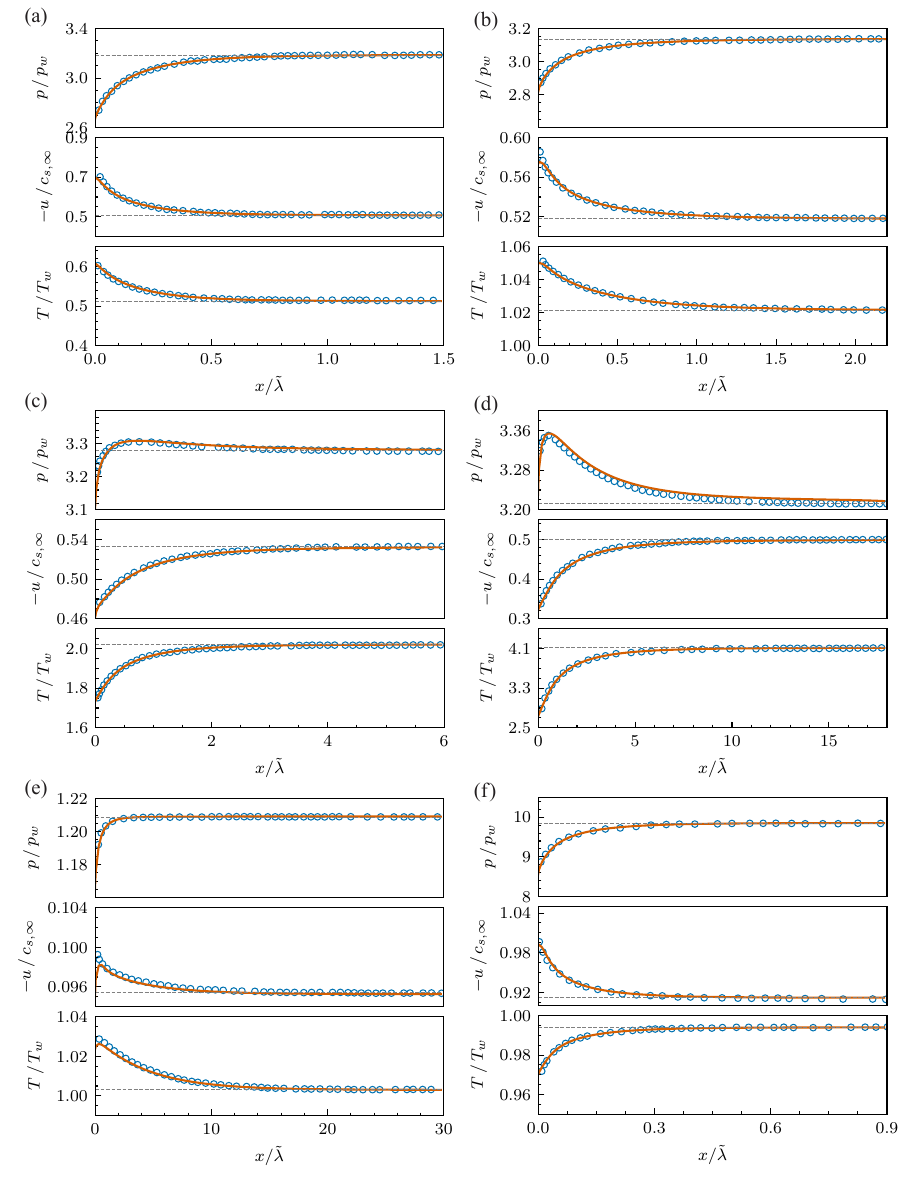}
    \vspace{-0.5em}
    \caption{
    Comparison of the present solution (solid lines) for the subsonic condensation problem with the reference \cite{Aoki1990} (circular markers).
    (a) $\MaNum_\infty =0.50757$, $p_\infty / p_w = 3.1858$, and $T_\infty/T_w =0.51220$;
    (b) $\MaNum_\infty = 0.51829$, $p_\infty/p_w = 3.1356$, and $T_\infty/T_w = 1.0213$;
    (c) $\MaNum_\infty = 0.53276$, $p_\infty/p_w = 3.2785$, and $T_\infty/T_w = 2.0194$;
    (d) $\MaNum_\infty = 0.50012$, $p_\infty/p_w = 3.2129$, and $T_\infty/T_w = 4.1124$;
    (e) $\MaNum_\infty = 0.095447$, $p_\infty/p_w = 1.2088$, and $T_\infty/T_w = 1.0030$;
    (f) $\MaNum_\infty = 0.91185$, $p_\infty/p_w = 9.8513$, and $T_\infty/T_w = 0.99389$.
    }
    \label{figAbsorbingCondensation}
\end{figure}

Fig.~\ref{figAbsorbingCondensation} compares $p/p_w$, $-u/c_{s,\infty}$, and $T/T_w$ as functions of $x/\tilde\lambda$.
The present curves capture both the approach to the far-field levels and the distinct near-wall shapes among the six configurations.
Cases (c) and (d) exhibit a near-wall pressure maximum before relaxing toward the far-field pressure, whereas the other panels show different pressure adjustments that are predominantly monotonic.
These comparisons support the reproduction of the principal condensation-layer profiles for the six listed far-field states, including the weak-inflow case and the configuration with the larger temperature ratio.

\subsection{Flow past a circular cylinder}

Flow past a diffuse-interface circular cylinder extends the tests to a curved boundary whose normal varies continuously relative to the Cartesian mesh and the discrete molecular velocities.
Low-speed cases assess integrated drag and wake geometry, while a high-speed rarefied case additionally probes strongly non-equilibrium gas dynamics and wall energy transfer.
For the low-speed tests, the Reynolds number $\ReNum=u_\infty D/\nu$ takes the values $20$ and $40$.
The domain is $35D\times25D$, the cylinder diameter is $D=1$, and its center is $(10D,10D)$.
Gas enters at the left boundary and exits at the right, with periodic top and bottom boundaries.
The prescribed values are $u_\infty=0.05$, $T_0=1.0$, and $\rho_0=1.0$, so that the cylinder radius is $R_c=D/2=0.5$.
The cylinder temperature is fixed to $T_0$.
The Mach number is below $0.1$, so comparison with incompressible benchmarks is appropriate.

\begin{figure}[!htp]
    \centering
    \includegraphics[width=0.8\textwidth]{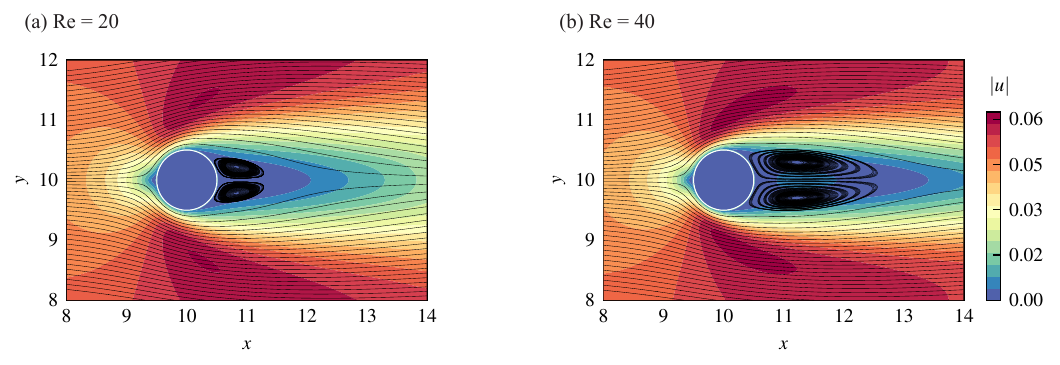}
    \caption{Steady velocity fields and streamlines at (a) $\rm Re = 20$ and (b) $\rm Re = 40$.
    }
    \label{figFlowAroundCylinderReStreamlines}
\end{figure}

An $8\times8$-point hGH quadrature is used in velocity space.
The AMR hierarchy starts from a $112\times64$ mesh, refines the interface and wake regions, and allows five successive refinements with ratio $2.0$.
The diffuse-interface width takes $\epsilon=3\Delta x_{\min}$ with $\Delta x_{\min}$ the finest mesh interval.
The Courant number is $C_{\rm CFL}=0.4$.
Fig.~\ref{figFlowAroundCylinderReStreamlines} shows an approximately symmetric recirculating wake at both Reynolds numbers and a longer wake at $\ReNum=40$.

The drag coefficient is defined by
\begin{equation}
    C_D=\frac{F_x}{0.5\rho_0 D u_\infty^2},
\end{equation}
where $F_x$ is the streamwise hydrodynamic force on the cylinder.
The recirculation length $L_w=2L_r/D$ is measured, where $L_r$ is a dimensional wake length.
Tab.~\ref{tab:FPC_ResultsCompare} provides a quantitative comparison with Refs.~\cite{LiuJ2022,NiuX2006}.
Using each reference value as the denominator, the DIKM drag at $\ReNum=20$ is higher by $1.38\%$ relative to Liu et al. and $2.47\%$ relative to Niu et al.; the corresponding differences at $\ReNum=40$ are $2.79\%$ and $4.34\%$.
The predicted wake lengths are lower than the reference values by at most $4.50\%$ at $\ReNum=20$ and $2.62\%$ at $\ReNum=40$.

\begin{table}[!htp]
    \centering
    \caption{Comparative results of the drag coefficient and dimensionless recirculation length between the DIKM and benchmarks.}
    \label{tab:FPC_ResultsCompare}
    \vspace{2pt}
    \renewcommand{\arraystretch}{1.1}
    \scalebox{0.85}{
    \begin{tabular}{*{7}{c}}
        \toprule
        \multirow{2}*{$\mathrm{Re}$} & \multicolumn{3}{c}{$C_D$} & \multicolumn{3}{c}{$2L_r/D$} \\
        \cmidrule(rl){2-4} \cmidrule(rl){5-7}
        & DIKM & Liu et al. \cite{LiuJ2022} & Niu et al. \cite{NiuX2006} 
        & DIKM & Liu et al. \cite{LiuJ2022} & Niu et al. \cite{NiuX2006} \\
        \midrule
        $20$ & $2.197$ & $2.167$ & $2.144$ & $1.805$ & $1.873$ & $1.890$ \\
        $40$ & $1.658$ & $1.613$ & $1.589$ & $4.421$ & $4.540$ & $4.520$\\
        \bottomrule
    \end{tabular}
    }
\end{table}

The high-speed calculation uses $\Omega=[-15R_c, 15R_c]^2$, a cylinder centered at $(0,0)$, with cylinder radius $R_c=1$.
The free-stream conditions are $\MaNum=5$, $\KnNum_\infty=\lambda_\infty/R_c=1$, $\rho_\infty = 1$ and $T_\infty=273\,\mathrm{K}$.
The stationary cylinder is maintained at $T_w=273\,\mathrm{K}$.
Argon enters from the left boundary, and zeroth-order extrapolation is applied at the other boundaries.
The VHS model is used, with the exponent $\omega=0.81$.
The AMR hierarchy starts from $48\times48$ cells and permits five refinements with ratio $2.0$.
The cylinder interface width is $\epsilon=3\Delta x_{\min}$ and $C_{\rm CFL}=0.4$.
The molecular-velocity discretization contains $4044$ points on an unstructured mesh, concentrated around the wall and free-stream distribution centers.

\begin{figure}[!htp]
    \centering
    \includegraphics[width=0.95\textwidth]{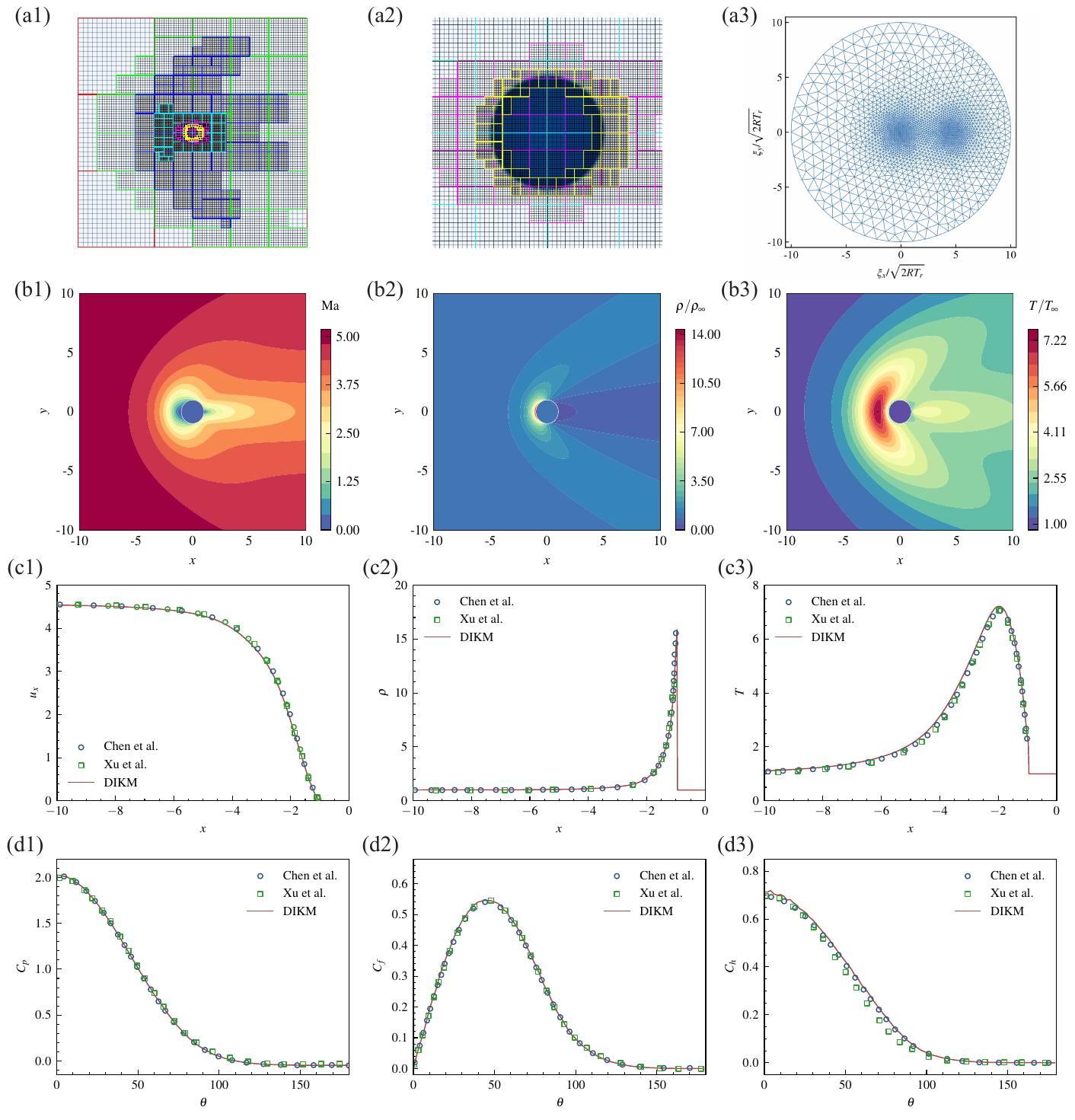}
    \caption{Discretization and simulation results of flow around a circular cylinder at $\MaNum=5$ and $\KnNum=1$.
    (a1) AMR block and mesh at steady state,
    (a2) the local zooming view near the circular cylinder,
    (a3) discretization of velocity space using non-structured mesh,
    (b1)-(b3) visualization of $x$-component velocity, density and temperature,
    (c1)-(c3) $x$-component velocity, density and temperature profiles along the stationary line,
    (d1)-(d3) pressure coefficient, heat transfer coefficient and friction coefficient distribution around the cylinder surface.
    }
    \label{figFlowAroundCylinderMa5Kn1}
\end{figure}

Fig.~\ref{figFlowAroundCylinderMa5Kn1}(a1)--(a2) shows the steady-state spatial mesh and its detail view around the cylinder, and (a3) the unstructured velocity mesh.
Fig.~\ref{figFlowAroundCylinderMa5Kn1}(b1)--(b3) draw the local Mach number, normalized density, and normalized temperature fields.
The upstream compression and heating extend over a resolved region ahead of the cylinder, and the wake differs markedly from the low-speed configurations.
Variables along the stagnation line $y=0$ are shown in Fig.~\ref{figFlowAroundCylinderMa5Kn1}(c1)--(c3).
The predicted streamwise velocity decrease, density rise near the front surface, and temperature maximum follow the reference profiles in Refs.~\cite{ChenJF2024CMA,XuD2026JCP}.

The quantities along the cylinder surface are examined, including the pressure coefficient $C_p$, friction coefficient $C_f$, and the heat flux coefficient $C_h$, defined as
\begin{equation}
    C_p=\frac{p_w-p_\infty}{0.5\rho_\infty u_\infty^2},\qquad
    C_f=\frac{\tau_w}{0.5\rho_\infty u_\infty^2},\qquad
    C_h=\frac{q_w}{0.5\rho_\infty u_\infty^3}.
\end{equation}
Here $p_w$, $\tau_w$, and $q_w$ denote the normal traction, tangential traction, and gas--wall heat flux, respectively.
For this case, the flow profiles in the bulk and the coefficients at the surface show that the DIKM works well for a stationary curved boundary in high-speed rarefied flow.

\subsection{Free motion of a piston in rarefied gas}

\begin{figure}[!htp]
    \centering
    \includegraphics[width=0.65\textwidth]{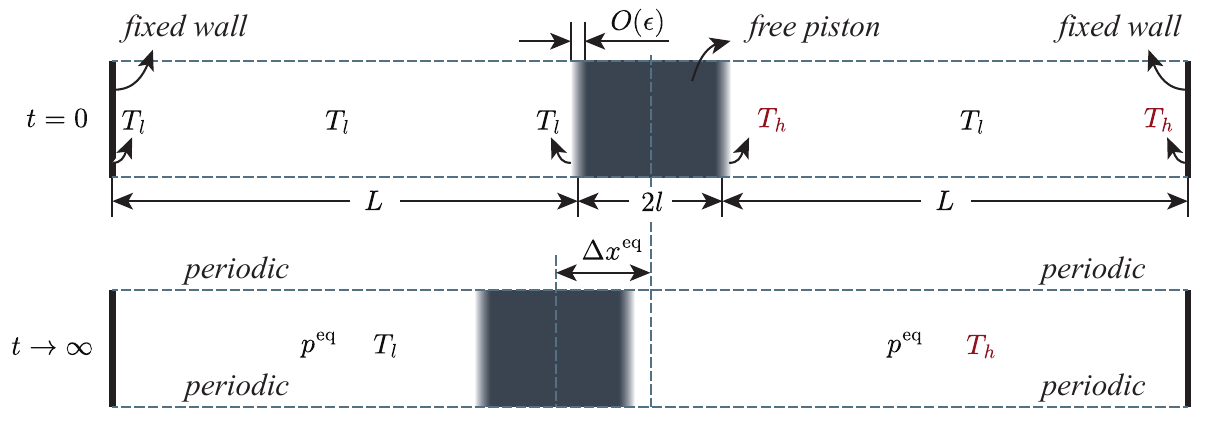}
    \caption{Schematic diagram of a free moving piston in rarefied gas.}
    \label{fig:piston_diagram}
\end{figure}

The freely moving piston tests feedback between molecular momentum transfer and the motion of a planar solid boundary.
Unlike a prescribed wall trajectory, the piston displacement depends on the time history of the computed gas forces.
The geometry in Fig.~\ref{fig:piston_diagram} has periodic top and bottom boundaries and fixed vertical walls at $T_l=270\,\mathrm{K}$ and $T_h=330\,\mathrm{K}$.
The piston has mass per unit area $M_c=3.56\times10^{-7}\,\mathrm{kg/m^2}$ and is initially centered at $x_c=0$.
The tube and piston lengths are $2(L+l)$ and $2l$, respectively, where $L=0.01\,\mathrm{m}$ and $l=0.001\,\mathrm{m}$.
The piston boundaries are described by diffuse interface, with width $\epsilon=3\Delta x$.
Both chambers initially contain initially stationary argon at $p_0=10\,\mathrm{Pa}$ and $T_0=T_l=270\,\mathrm{K}$.
The left and right piston faces are held at $T_l$ and $T_h$, respectively.
The stated Knudsen number is $0.31$, referenced to the piston length, following the Refs.~\cite{HeQ2024IJHMT,Shrestha2015JCP}.

Heating of the right chamber induces pressure imbalance that drives the piston leftward.
Each chamber retains its initial gas mass and approaches a uniform temperature equal to its two bounding surfaces, pressure equality gives an equilibrium displacement magnitude
\[
    \Delta x^{\rm eq}=L\frac{T_h-T_l}{T_l+T_h}.
\]
The corresponding final piston position is $x_c^{\rm eq}=-\Delta x^{\rm eq}=-1\,\mathrm{mm}$.
The calculation uses a uniform $440\times2$ mesh and a $49\times49$-point Newton--Cotes (NC) velocity quadrature over $[-5\sqrt{2RT_0},5\sqrt{2RT_0}]$.
The Courant number uses $C_{\rm CFL} = 0.4$.

\begin{figure}[!htp]
    \centering
    \includegraphics[width=0.8\textwidth]{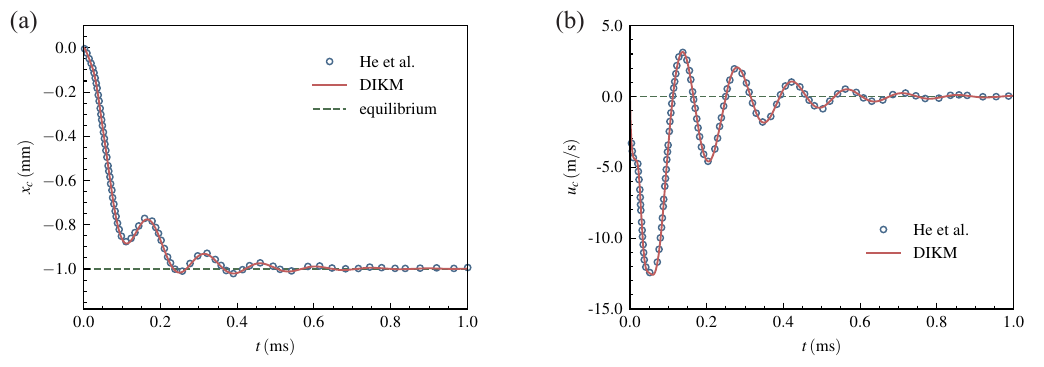}
    \caption{The variation of (a) piston center position and (b) velocity with respect to time.}
    \label{fig:piston_vel_pos}
\end{figure}

Fig.~\ref{fig:piston_vel_pos} compares piston position and velocity with the immersed-boundary DUGKS results of He et al.~\cite{HeQ2024IJHMT}.
The piston first accelerates toward the cold side, then undergoes damped reversals while its position approaches the analytical equilibrium and its velocity approaches zero.
The present curves follow the principal oscillations and the decay envelope of the reference markers over the displayed interval.
A minimum velocity of approximately $-1.26\,\mathrm{mm/ms}$ occurs at $0.0589\,\mathrm{ms}$ and a subsequent zero crossing at $0.112\,\mathrm{ms}$.
Comparisons of the equilibrium position and transient response support the model's ability to capture the main features of coupled translational dynamics.

\subsection{Free motion of a particle in a micro lid-driven cavity}

\begin{figure}[!htp]
    \centering
    \includegraphics[width=0.8\textwidth]{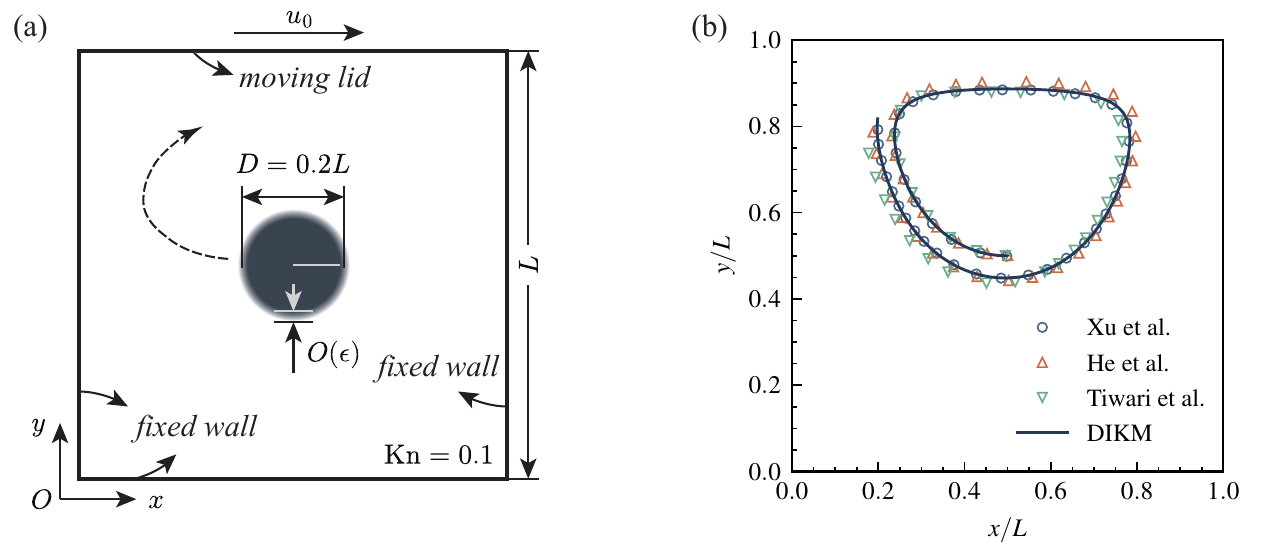}
    \caption{Simulation case of particle moving in micro cavity flow. (a) Schematic diagram of computational setup, (b) particle center trajectory history compared with benchmarks \cite{Tiwari2022JCP,XuD2026JCP,HeQ2024IJHMT}.}
    \label{fig:Comb_diag_trajectory}
\end{figure}

A freely moving circular particle in a lid-driven cavity combines a curved boundary, translation, rotation, and transient rarefied flow.
The schematic diagram for this case is shown in Fig.~\ref{fig:Comb_diag_trajectory} (a).
The particle initially lies at the center of a square cavity with side length $L=10^{-6}\,\mathrm{m}$ and diameter $D=L/5$.
Its density is $\rho_s$, and the initial gas density satisfies $\rho_0=\rho_s/10 = 1.0$.
The gas is argon with initial temperature $T_0=270\,\mathrm{K}$, and the cavity walls and particle surface are maintained at $T_w = T_s = 270\,\mathrm{K}$.
The top lid moves rightward at $u_0=30\,\mathrm{m/s}$, while the other cavity walls remain fixed.
The Knudsen number is $\KnNum=\lambda/L=0.1$.

The velocity discretization uses a $41\times41$-point NC quadrature over $[-5\sqrt{2RT_0},5\sqrt{2RT_0}]$, and the Courant number is $C_{\rm CFL}=0.3$.
The AMR hierarchy starts from $64\times64$ cells and permits two refinements capturing interface region, with $L_M = 3$ and $\epsilon=3\Delta x_{\min}$.
The particle trajectory is plotted in Fig.~\ref{fig:Comb_diag_trajectory} (b), compared with Refs.~\cite{Tiwari2022JCP,HeQ2024IJHMT, XuD2026JCP}.
The particle moves clockwise and follows the reference trajectories well as the lid-driven circulation transports it through the cavity.

\begin{figure}[!htp]
    \centering
    \includegraphics[width=\textwidth]{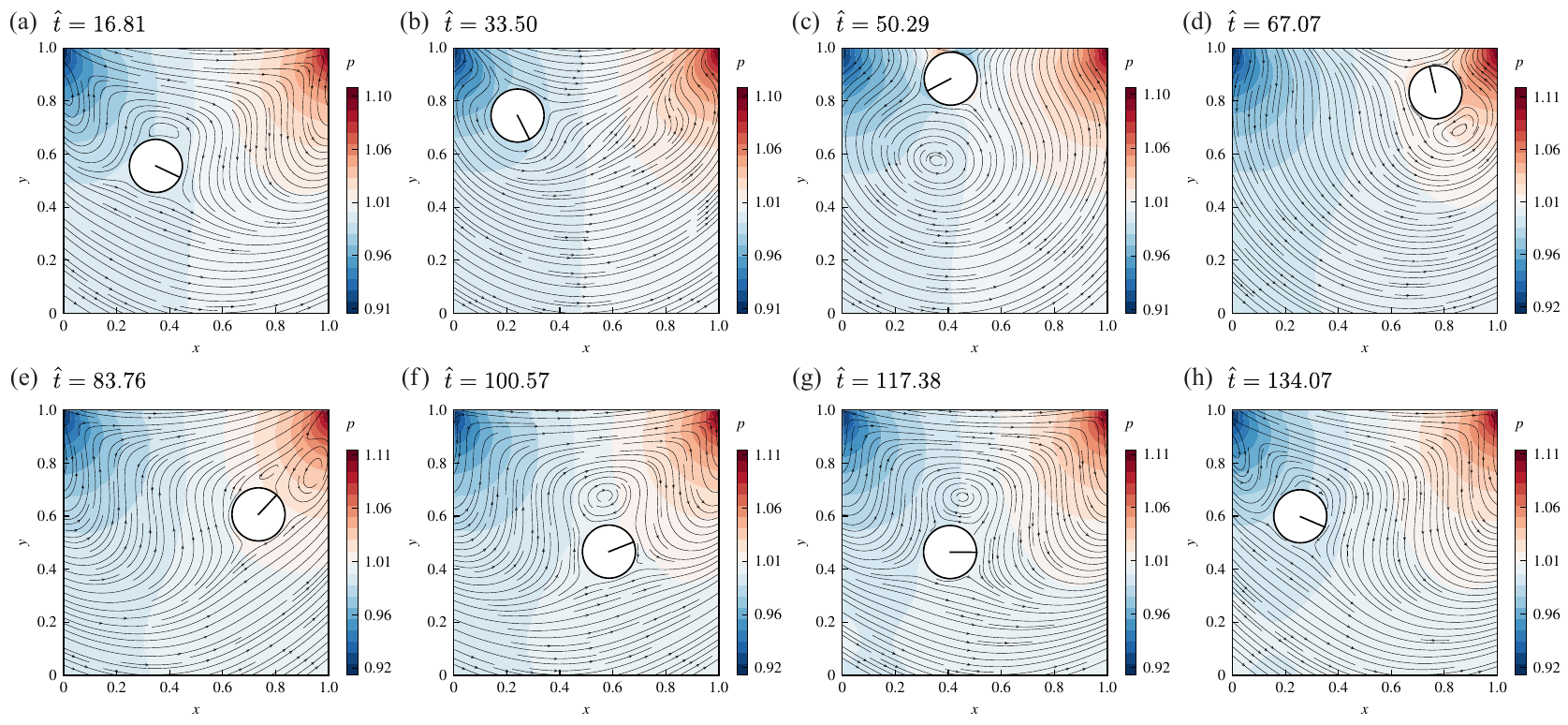}
    \caption{Normalized pressure fields $p/\rho_0 RT_0$ and streamlines captured at $\hat t = 16.81$, $33.50$, $50.29$, $67.07$, $83.76$, $100.57$, $117.38$ and $134.07$, respectively.
    Streamlines are plotted in a coordinate system moving with the particle.
    }
    \label{fig:Comb_Press_Streamlines}
\end{figure}

Pressure fields and streamlines are presented in Fig.~\ref{fig:Comb_Press_Streamlines}, at selected dimensionless times $\hat{t}=t/t_r=16.81$, $33.50$, $50.29$, $67.07$, $83.76$, $100.57$, $117.38$, and $134.07$, where $t_r=L/\sqrt{2RT_0}$.
The pressure fields are normalized by $\rho_0 RT_0$.
The particle undergoes clockwise rotation driven by the shear induced by the cavity flow.
The minimum and maximum pressures occur near the upper-left and upper-right corners of the cavity, respectively.
As the particle approaches the moving lid, a pronounced horizontal pressure gradient develops in the narrow gap, accelerating both the translational and rotational motion of the particle.
The predicted pressure distributions, streamlines, particle positions, and orientations agree well with those reported by Xu et al. \cite{XuD2026JCP} using the immersed boundary method, demonstrating the capability of the DIKM to accurately capture the translational and rotational motion of a solid particle in rarefied gas flow.

\subsection{Physical vapor deposition in a rarefied gas}

The final example examines whether the DIKM can couple rarefied vapor transport to absorption and the subsequent evolution of a material surface.
The vapor is assumed monatomic and single-component, and the material surface energy is isotropic.
The substrate temperature and absorption probability are fixed, and evaporation, nucleation, elastic stress, and latent-heat feedback are omitted.

The vapor has reference temperature $T_0 = 1$, density $\rho_0 = 1$, and $p_0=\rho_0 RT_0$, with $R=0.5$.
The reference length $L=1$.
The incident vapor velocity is $\bm u_0=(\sin\alpha,-\cos\alpha)$, with unit magnitude and incidence angle $\alpha$ measured from the downward vertical direction.
The Prandtl number is $\PrNum=2/3$ and the VHS viscosity law uses $\omega=0.81$.
Knudsen number is $\KnNum = 1$, specified relative to reference length.
Velocity space uses a tensor-product $32\times32$-point hGH quadrature.

Let $\varphi=1-\phi$ denote the thin-film order parameter, with $\varphi=1$ in the film and $\varphi=0$ in the vapor. Following the phase-field description of deposition and surface diffusion \cite{Salvalaglio2016PRB}, we use total free energy functional of the system
\begin{equation}
    \calF (\varphi) = \int_\Omega \left[
        \dfrac{18}{\epsilon} \varphi^2(1-\varphi)^2 
        + \dfrac{\epsilon}{2}|\nabla\varphi|^2
    \right] \diff V.
\end{equation}
The evolution equation is a Cahn--Hilliard equation with degenerate mobility and a deposition source term
\begin{equation}
    \dfrac{\partial \varphi}{\partial t}
    =  
    \nabla \cdot [M(\varphi) \nabla \mu]
    + S_\varphi
    ,
    \label{eqCHEqFilmGrowth}
\end{equation}
where $\mu$ is the chemical potential and $S_\varphi = R_{\rm abs} / \rho_s$ accounts for the growth of the film.
The degenerate mobility $M(\varphi)=(36M_0/\epsilon)\varphi^2(1-\varphi)^2$ vanishes in bulk phases and localizes diffusive transport to the interfacial region.
The constant $M_0$ controls an effective surface diffusivity rate.
The chemical potential is obtained from the variation of the total free energy
\begin{equation}
    \label{eqFilmGrowthChemPotent}
    \mu = \dfrac{\delta \calF}{\delta \varphi} = 
     \dfrac{36}{\epsilon} \varphi (1 - \varphi) (1 - 2\varphi)
     - \epsilon \nabla^2 \varphi.
\end{equation}
Here $R_{\mathrm{abs}}$ is the gas mass removed per unit volume and time by the DIKM absorption update.
Equation~\eqref{eqCHEqFilmGrowth} is discretized using a finite-volume scheme, with second-order central differences and arithmetic averaging of mobility at cell faces.
Time integration uses the explicit second-order Heun method, with the deposition source held fixed over a time step.

\begin{figure}[!htp]
    \centering
    \includegraphics[width=\linewidth]{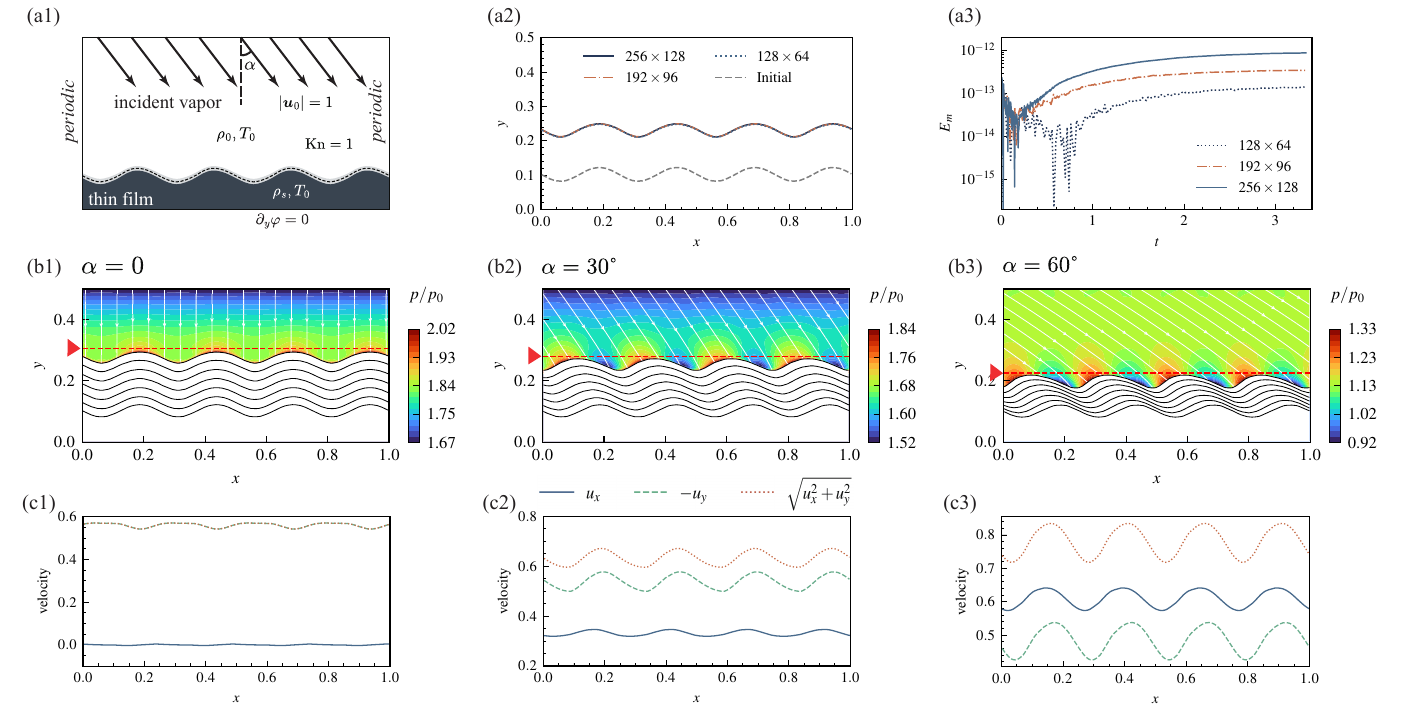}
    \caption{
    Vapor deposition and film evolution at $\KnNum=1$.
    (a1) Computational configuration.
    (a2) Film-height profiles at $t=3$ for three uniform grids. Gray dash-dotted curve denotes the initial profile.
    (a3) Normalized total-mass residual $E_M$ versus time.
    (b1)--(b3) Normalized pressure $p/p_0$ (colors) and gas streamlines (white) at $t=4$, with successive film profiles (black), for $\alpha=0^\circ$, $30^\circ$, and $60^\circ$, respectively.
    (c1)--(c3) Gas velocity components $u_x$ (blue solid), $-u_y$ (green dashed), and speed $\sqrt{u_x^2+u_y^2}$ (red dotted), sampled along the red horizontal dashed lines in the corresponding panels (b1)--(b3) at $t=4$.}
    \label{fig:pvd-growth}
\end{figure}

The domain is $[0, L]\times[0, L/2]$, shown in 
Fig.~\ref{fig:pvd-growth}(a1).
The mobility scale is set to $M_0 = 1\times 10^{-8}$.
The density ratio is set to $\rho_s/\rho_0=20$.
The vapor is initialized at equilibrium with $\rho_0, T_0$, and $\bm u_{0}$.
The interface width is $\epsilon = 3\Delta x$, and Courant number $C_{\rm CFL} = 0.4$.
The initial film is specified by
\begin{equation}
    \varphi(x,y,0)=\frac12\left[1-\tanh\left(
    \frac{3[y-0.1 + 0.02\sin(8\pi x)]}{0.04}\right)\right].
    \label{eq:pvd-initial}
\end{equation}
The lateral boundaries are periodic for both fields.
At the top, the reduced distributions entering the domain are taken as the equilibrium distribution with $\rho_0, T_0$, and $\bm u_{0}$.
The phase field satisfies $\partial_n\varphi=\partial_n\mu=0$ at both horizontal boundaries. The absorption probability is $\beta = 0.8$.
First, the grid independence and total mass conservation is tested.
The gas and film mass are $M_g=\int_\Omega\phi\rho\,\mathrm dV$ and $M_s=\rho_s\int_\Omega\varphi\,\mathrm dV$, and mass rates entering and leaving through the open boundaries are $\dot M_{\mathrm{in}}$ and $\dot M_{\mathrm{out}}$.
The total-mass residual is thus defined by
\begin{equation}
  R_M=M_g(t)+M_s(t)-M_g(0)-M_s(0)
    -\int_0^t(\dot M_{\mathrm{in}}-\dot M_{\mathrm{out}})\,\mathrm dt'.
  \label{eq:pvd-mass}
\end{equation}
The relative mass conservation error is defined as the normalized residual
\begin{equation}
    E_M(t)=\frac{|
    R_M(t)
    |}{M_{\mathrm{abs}}(t)},
\end{equation}
where
$M_{\mathrm{abs}}(t)=\int_0^t\int_\Omega
R_{\mathrm{abs}}(\bm x,t')\,\mathrm dV\,\mathrm dt'$
is the cumulative absorbed mass.
For the grid convergence test, $\alpha=0$ and three uniform meshes of $256\times128$, $192\times96$, and $128\times64$ cells are used, without adaptive refinement.
The film-height profiles at $t=3$ nearly overlap in Fig.~\ref{fig:pvd-growth}(a2), indicating weak sensitivity to the tested spatial resolutions.
Fig.~\ref{fig:pvd-growth}(a3) shows normalized total-mass residuals of order $10^{-12}$ or smaller over the plotted interval.
It demonstrates that discrete mass conservation of the coupled absorption, deposition, and boundary transport is essentially satisfied for this test.

The effect of incidence angle is then examined at fixed incident speed using a $192\times96$ mesh, and $\alpha=0^\circ$, $30^\circ$, and $60^\circ$.
No physical or numerical perturbation is introduced.
The successive profiles in Fig.~\ref{fig:pvd-growth}(b1)--(b3) show sustained deposition in all three cases, with final time $t=4$.
Normal incidence approximately preserves the symmetry of the film surface, whereas oblique incidence produces increasingly asymmetric profiles, most clearly at $60^\circ$.
The final film height decreases as the incidence angle increases.
The pressure fields exhibit an increase toward the film and a spatial modulation following the surface corrugations.
The pressure is higher on the wave-facing flank, while a localized low-pressure region forms on the lee flank near the trough.
The streamlines become more inclined with increasing $\alpha$ and deflect locally near the corrugated surface.
Panels (c1)--(c3) quantify the gas velocity along the corresponding horizontal sampling lines above the film.
At normal incidence, $u_x$ remains close to zero, while $-u_y$ and the speed nearly coincide.
At oblique incidence, a positive streamwise component develops and accounts for an increasing fraction of the speed, while all three quantities retain a periodic modulation associated with the surface geometry.
The results demonstrate the capability of the DIKM to realize coupling between vapor transport and surface evolution.

\section{Conclusions}
\label{secConclusion}

This work develops a flux-balanced diffuse-interface kinetic model for rarefied gas--solid interactions on a fixed Eulerian mesh.
A localized boundary-flux operator balances incident removal, prescribed absorption, and diffuse reflection, with its velocity moments determining interfacial mass, momentum, and energy exchange.
The discrete source update preserves this balance by returning the reflected fraction of the mass actually removed.
Matched asymptotic analysis recovers the prescribed absorption--diffuse-reflection law in the sharp-interface limit for smooth, kinematically compatible stationary or moving interfaces.
The formulation thus connects implicit geometry to molecular boundary physics while recovering the original kinetic equation in the bulk gas.

Benchmark comparisons show that the model reproduces changes in velocity slip and thermal transport with rarefaction across the tested planar and curved geometries.
Absorbing-boundary tests verified recovery of the prescribed molecular loss and the associated non-equilibrium flow.
Moving-body comparisons further demonstrate that the interfacial momentum exchange captures gas-driven translation and rotation.
An illustrative thin-film growth case couples adsorbed vapor to material accumulation and exhibits distinct morphological responses at different vapor incident angles.
This example provides preliminary evidence that the model can be extended to study interactions between evolving material interfaces and rarefied gases.
By organizing gas--solid coupling around molecular flux balances, the DIKM provides a common basis for linking rarefied transport and material interface evolution.

\section{Acknowledgement}
The authors would like to thank Professor Zhenhua Chai (Huazhong University of Science and Technology) for his valuable discussions and constructive suggestions.
This work was supported by the National Key Research and Development Program of China (No. 2023YFB3710202).

\appendix
\section{Derivation of diffuse-interface kinetic model}

\label{secAppendixA}

This appendix derives Eq.~\eqref{eqDIKM} for a gas domain $\Omega_g(t)$ bounded by a moving gas--solid interface $\Gamma(t)$.
Inside $\Omega_g(t)$, the distribution function satisfies the Shakhov--Boltzmann equation
\begin{equation}
    \partial_t f+\bm \xi\cdot\nabla f=Q^{\rm S}(f),
    \qquad \bm x\in\Omega_g(t).
    \label{eq:app_sharp_boltzmann_moving}
\end{equation}
The collision term used in this study is the Shakhov model.
Nevertheless, the DIKM is not constrained to any specific type of collision model.
The boundary condition to close the problem is defined on $\Gamma(t)$, specified as partially absorbing diffuse-reflection boundary with absorption probability $\beta\in[0,1]$.
The sharp-interface distribution, with the emitted half-space populated by the unit-density wall Maxwellian defined in Eq.~\eqref{eq:wall_maxwellian}, can be written as
\begin{equation}
    f_\Gamma
    =\chi^-(c_n)f^-
    +\chi^+(c_n)\rho_\Gamma\widehat f_\Gamma^{\rm M},
    \label{eq:app_sharp_boundary_distribution}
\end{equation}
The reflected density $\rho_\Gamma$ is determined by the local normal-flux balance. If a fraction $\beta$ of the incident molecular flux is absorbed and the remaining fraction is diffusely reflected, then
\begin{equation}
    \rho_\Gamma
    =
    \frac{
        (1-\beta)
        \displaystyle\int_{c_n<0}|c_n|f^-\diff\bm \xi
    }{
        \displaystyle\int_{c_n>0}c_n\widehat f_\Gamma^{\rm M}\diff\bm \xi
    }.
    \label{eq:app_rho_gamma_flux_balance}
\end{equation}

Let $\psi(\bm x,t)$ be a sufficiently smooth test function, compactly supported in the fixed computational domain $\Omega$ in space. Multiplying Eq.~\eqref{eq:app_sharp_boltzmann_moving} by $\psi$ and integrating over the time-dependent gas domain for $0<t<T$ gives
\begin{equation}
    \label{eq:app_space_time_weak_start}
    \begin{aligned}
        &\int_0^T\int_{\Omega_g(t)}
            \psi\,\partial_t f\diff\bm x\diff t
        +\int_0^T\int_{\Omega_g(t)}
            \psi\,\bm \xi\cdot\nabla f\diff\bm x\diff t
        \\
        &\hspace{4cm}=
        \int_0^T\int_{\Omega_g(t)}
            \psi Q^{\rm S}(f)\diff\bm x\diff t.
    \end{aligned}
\end{equation}

With the help of the Reynolds transport theorem, we have
\begin{equation}
    \label{eq:app_time_integration_moving}
    \begin{aligned}
        \int_0^T\int_{\Omega_g(t)}\psi\partial_t f\diff\bm x\diff t
        ={}&
        -\int_0^T\int_{\Omega_g(t)}f\partial_t\psi\diff\bm x\diff t
        +\int_0^T\int_{\Gamma(t)}\psi fV_{n,\Gamma}\diff s\diff t
        \\
        &+\int_{\Omega_g(T)}\psi(\bm x,T)f(\bm x,T)\diff\bm x
        -\int_{\Omega_g(0)}\psi(\bm x,0)f(\bm x,0)\diff\bm x.
    \end{aligned}
\end{equation}
The term containing $V_{n,\Gamma}$ in Eq.~\eqref{eq:app_time_integration_moving} is absent for a stationary boundary and is the key modification introduced by domain motion.
Because $\bm \xi$ is independent of $\bm x$, one has $\bm \xi\cdot\nabla f=\nabla\cdot(\bm \xi f)$. Performing integration by parts in $\Omega_g(t)$ gives
\begin{equation}
    \label{eq:app_convection_integration_moving}
    \begin{aligned}
        \int_{\Omega_g(t)}\psi\bm \xi\cdot\nabla f\diff\bm x
        =-\int_{\Omega_g(t)}f\bm \xi\cdot\nabla\psi\diff\bm x
        -\int_{\Gamma(t)}\psi f\bm \xi\cdot\bm n_{\Gamma}\diff s.
    \end{aligned}
\end{equation}

Substituting Eqs.~\eqref{eq:app_time_integration_moving} and \eqref{eq:app_convection_integration_moving} into Eq.~\eqref{eq:app_space_time_weak_start} gives the space--time weak form
\begin{equation}
    \label{eq:app_full_weak_moving}
    \begin{aligned}
        &-\int_0^T\int_{\Omega_g(t)}
            f\left(\partial_t\psi+\bm \xi\cdot\nabla\psi\right)\diff\bm x\diff t
        -\int_0^T\int_{\Gamma(t)}\psi c_n f_\Gamma\diff s\diff t
        \\
        &\quad
        +\int_{\Omega_g(T)}\psi(\bm x,T)f(\bm x,T)\diff\bm x
        -\int_{\Omega_g(0)}\psi(\bm x,0)f(\bm x,0)\diff\bm x
        \\
        &\hspace{3cm}=
        \int_0^T\int_{\Omega_g(t)}\psi Q^{\rm S}(f)\diff\bm x\diff t.
    \end{aligned}
\end{equation}

Introduce the characteristic function of the moving gas domain
\begin{equation}
    \chi_g(\bm x,t)=
    \begin{cases}
        1, & \bm x\in\Omega_g(t),\\
        0, & \bm x\in\Omega\setminus\Omega_g(t).
    \end{cases}
    \label{eq:app_characteristic_moving}
\end{equation}
Eq.~\eqref{eq:app_full_weak_moving} is then extended to the larger domain $\Omega$
\begin{equation}
    \begin{aligned}
        - & \int_{0}^{T} \int_{\Omega} \chi_{g} f\left(\partial_{t} \psi+\bm{\xi} \cdot \nabla \psi\right) \diff \bm x \diff t 
        -\int_{0}^{T} \int_{\Omega} \delta_{\Gamma} \psi c_{n} f_{\Gamma} \diff \bm x \diff t \\
        + & \int_{\Omega} \chi_{g}(\bm x, T) \psi(\bm x, T) f(\bm x, T) \diff \bm x 
        -\int_{\Omega} \chi_{g}(\bm x, 0) \psi(\bm x, 0) f(\bm x, 0) \diff \bm x \\
        = & \int_{0}^{T} \int_{\Omega} \chi_{g} \psi Q^{\mathrm{S}}(f) \diff \bm x \diff t
    \end{aligned}
\end{equation}

Using integration by parts, we obtain the weak form formulation
\begin{equation}
    \int_{0}^{T} \int_{\Omega} \psi \left[\partial_{t}\left(\chi_{g} f\right)+\bm{\xi} \cdot \nabla\left(\chi_{g} f\right)
    -\chi_{g} Q^{\mathrm{S}}(f)-\delta_{\Gamma} c_{n} f_{\Gamma}\right] \diff \bm{x} \diff t=0
    ,
\end{equation}
and it leads to
\begin{equation}
    \partial_t(\chi_{g} f)
    +\bm \xi\cdot\nabla(\chi_{g}f)
    =\chi_{g}Q^{\rm S}(f)
    +\delta_\Gamma c_n f_\Gamma.
    \label{eq:app_sharp_distribution_moving}
\end{equation}

Replacing the characteristic function $\chi_{g}$ by the smooth gas order parameter $\phi\in[0,1]$ introduced in Section~\ref{subsec:diffuse_regularization}, the surface Dirac measure $\delta_\Gamma$ by its regularized counterpart $\delta_\epsilon=|\nabla\phi|$, and $\rho_\Gamma$ by its diffuse-interface counterpart $\rho_\epsilon$ gives the diffuse-interface kinetic model (Eqs.~\eqref{eqDIKM}, \eqref{eq:local_boundary_operator} and \eqref{eq:rho_gamma_local}).



\bibliographystyle{cas-model2-names}



\end{document}